\documentclass[%
final,
nofootinbib,
amsmath,amssymb,
aps,
floatfix,
]{revtex4-2}

\usepackage{lineno}
\usepackage{graphicx}% Include figure files
\usepackage{dcolumn}% Align table columns on decimal point
\usepackage{bm}% bold math
\usepackage[noabbrev,capitalise]{cleveref}
\crefname{equation}{equation}{equations}
\Crefname{equation}{Equation}{Equations}
\usepackage{xcolor}
\usepackage{appendix}
\usepackage{subcaption}
\usepackage[linesnumbered,ruled]{algorithm2e}

\usepackage{soul}
\setstcolor{blue}

\usepackage{tikz}
\usetikzlibrary{shapes,arrows}

\newcommand{\J}{\mathcal{J}}
\newcommand{\avgJ}{\langle\mathcal{J}\rangle}
\newcommand{\avg}[1]{\left\langle #1 \right\rangle}
\newcommand{\W}{\bm W}
\newcommand{\subin}{\mathrm{in}}
\newcommand{\subout}{\mathrm{out}}
\newcommand{\Wout}{\W_\subout}
\newcommand{\Win}{\W_\subin}
\newcommand{\dpartial}[2]{\ensuremath{\frac{\partial{#1}}{\partial{#2}}}}
\newcommand{\Eac}{E_\mathrm{ac}}
\newcommand{\R}{\mathbb{R}}
\newcommand{\K}{\bm{\mathcal{K}}}

\newcommand{\fh}{\textcolor{black}}
\soulregister{\fh}{1}
\newcommand{\Sref}[1]{Section \ref{#1}}

\begin{document}

\preprint{APS/123-QED}

\title{Gradient-free optimization of chaotic acoustics with reservoir computing}% Force line breaks with \\

\author{Francisco Huhn}
\affiliation{%
University of Cambridge, Department of Engineering, United Kingdom
}%
\author{Luca Magri}%
\email{lm547@cam.ac.uk}
\affiliation{%
Imperial College London, Aeronautics Department, United Kingdom
}%
\affiliation{%
University of Cambridge, Department of Engineering, United Kingdom
}%
\affiliation{%
The Alan Turing Institute, United Kingdom
}%
\affiliation{%
Institute of Advanced Study, TU Munich, Germany (visiting)
}%

\date{\today}% It is always \today, today,

\begin{abstract}
Gradient-based optimization of chaotic acoustics is challenging for a threefold reason: (i) first-order perturbations grow exponentially in time; (ii) the statistics of the solution may have a slow convergence; (iii) and the time-averaged acoustic energy may physically have discontinuous variations, which means that the gradient does not exist for some design parameters.
We develop a versatile optimization method, which finds the design parameters that minimize time-averaged acoustic cost functionals, and overcomes the three aforementioned challenges. The method is gradient-free, model-informed, and data-driven with  reservoir computing based on echo state networks. 
First, we analyse the predictive capabilities of echo state networks in thermoacoustics both in the short- and long-time prediction of the dynamics. We find that both fully data-driven and model-informed architectures are able to learn the chaotic acoustic dynamics, both time-accurately and statistically. 
Informing the training with a physical reduced-order model with one acoustic mode markedly improves the accuracy and robustness of the echo state networks, while keeping the computational cost low. 
Echo state networks offer accurate predictions of the long-time dynamics, which would be otherwise expensive by integrating the governing equations to evaluate  the time-averaged quantity to optimize. 
Second, we couple echo state networks with a Bayesian technique to explore the design thermoacoustic parameter space.  The computational method is minimally intrusive because it requires only the initialization of the physical and hyperparameter optimizers. 
Third, we find the set of flame parameters that minimize the time-averaged acoustic energy of chaotic  oscillations, which are caused by the positive feedback with a heat source, such as a flame in gas turbines or rocket motors. These oscillations are known as thermoacoustic oscillations.  
The optimal set of flame parameters is found with the same accuracy as brute-force grid search, but with a convergence rate that is more than one order of magnitude faster. 
This work opens up new possibilities for non-intrusive (``hands-off'') optimization of chaotic systems, in which the cost of generating data, for example from high-fidelity simulations and experiments, is high. 

\end{abstract}

\maketitle

\section{Introduction}
\label{sec:intro}

% 1. TA is a problem that we care about
When the heat released by a flame is sufficiently in phase with the acoustic waves of a confined environment, such as a gas turbine, thermoacoustic oscillations can arise~\cite{Rayleigh1878}. 
Physically, thermoacoustic oscillations occur when the thermal power released by the flame, which is converted into acoustic energy, exceeds fluid mechanic dissipation.
In gas turbines and rocket motors, thermoacoustic oscillations are unwanted because they can  cause structural vibrations, fatigue, noise, and, if uncontrolled, can shake the device apart.
Therefore, the objective of manufacturers is to design and operate stable devices~\citep{Lieuwen2005,Culick2006,Dowling2015}. %,Poinsot2017,Juniper2018}. 
%
% 2. TA needs to be eliminated by optimization.  Practice is linear methods. 
%
The preliminary design of thermoacoustic systems is based on linear analysis, in which the growth rates of infinitesimal oscillations is computed on top of a time-independent baseline solution. 
If no growth rate is positive, the system is linearly stable~\citep{Dowling1997,Lieuwen2005,Juniper2018,Magri2018}. 
If the system is linearly unstable, sensitivity methods, which are based on  gradient computation, have been introduced to answer the practitioners' question 
``How can we change the design parameters to reduce the growth rate of infinitesimal oscillations?''. 
In particular, adjoint methods proved computationally cheap tools to calculate the gradients of an eigenvalue with respect to all parameters of interest~\citep{Magri2013,Magri2014,Orchini2015b,Mensah2017,Magri2018} with higher-order corrections~\citep{Mensah2018}. 
Adjoint gradients were applied to the optimization of a longitudinal combustor~\citep{Aguilar2020} and annular combustors~\citep[e.g.,][]{Aguilar2018,Schaefer2021}. 

% 3. But TA is nonlinear 
Although linear analysis provides valuable information on the system's stability and sensitivity, thermoacoustic oscillations are inherently nonlinear.
First, the heat release varies nonlinearly with the acoustics, which perturb the flame dynamics~\citep{Dowling1997,Dowling1999}. 
Second, hydrodynamic instabilities (e.g. vortex shedding), which are promoted by the geometry of the combustor, can modulate the flame dynamics and, thus, the heat release rate~\cite{Huhn2020b}.
Because of these nonlinearities, a thermoacoustic system may be stable to infinitesimal perturbations (i.e. linearly stable),
but finite-amplitude perturbations can trigger sustained oscillations~\citep{Subramanian2011}, i.e. in the bistable region of a subcritical bifurcation.
These sustained oscillations can be periodic, quasi-periodic or chaotic, as shown in experiments~\citep{Kabiraj2011,Gotoda2011,Gotoda2012,Kabiraj2012,Nair2014,Nair2015} and numerical studies~\citep{Kashinath2014,Waugh2014,Orchini2015}, to name a few. 
Among these nonlinear regimes, chaotic oscillations are the most intractable to optimization~\citep{Huhn2020a,Huhn2020b}. 

% 4. Chaotic solutions. Eigenvalue gradient methods cannot optimize chaos.
Chaotic oscillations are extremely sensitive to infinitesimal perturbations~\citep{Lorenz1963},  
which results in an exponential growth of infinitesimal perturbations.
%, which means that,
In other words, the tangent space in unstable\footnote{In other words, at least one Lyapunov exponent is positive.}. 
Because of this, the calculation of gradients of ergodic averages, i.e. {\it time-averaged} quantities of interest, is intractable with traditional sensitivity methods. 
%Although it is possible to extend traditional eigenvalue analysis to compute the sensitivity of periodic orbits with Floquet theory~\citep{HuhnJFM}, eigenvalue analysis cannot be applied to chaotic thermoacoustics~\citep{Huhn2020a,Huhn2020b}. 
This roadblock motivated the development of alternative gradient-based methods, which can be grouped into six categories: 
(i) ensemble methods \citep{Lea2000,Lea2002,Eyink2004}, %\lm{which ... }
which average the gradient over an ensemble of short time trajectories; 
(ii) probability density methods \citep{Thuburn2005,Blonigan2014}, %\lm{which ... }
which calculate the gradient from the change in the probability density function of the chaotic attractor; 
(iii) unstable periodic orbits \citep{Lasagna2018}, %\lm{which ... }
which decompose the chaotic attractor into unstable periodic orbits and compute their gradients; 
(iv) fluctuation-dissipation-theorem methods \citep{Leith1975,Abramov2007,Abramov2008}, which compute the mean linear response of a system to small changes in external forcing;
(v) shadowing methods \citep{Wang2013,Wang2014,Ni2017,Blonigan2018,Ni2019a},
which average over time the difference between a baseline trajectory and its shadowing trajectory, and (vi) recent developments on linear response theory~\citep{chandramoorthy2020computable,ni2020fast}.
%
% 6. Shadowing methods 
%
In particular, shadowing methods have  successfully computed  first-order sensitivities of time-averaged energies in fluid mechanics~%\lm{[... mention fluids areas of applications, but not TA.]} \fh{???}
\citep{Blonigan2017,Ni2019b}. 
In thermoacoustics, shadowing-based gradients were embedded into a gradient update routine for design optimization~\citep{Huhn2020a}, in which the time-averaged acoustic energy was minimized by computing the optimal set of flame parameters. 
%\lm{[The authors investigated a thermoacoustic case. Main findings, success and conclusions.]}
The study highlighted three challenges in gradient-based optimization of chaotic thermoacoustics. 
First, thermoacoustic systems physically exhibit an abundance of bifurcations, across which the time-averaged cost functional being optimized can be discontinuous~\citep{Huhn2020a,Huhn2020b}.
Second, shadowing-based methods require a number of tangent (i.e. first-order perturbation) solutions equal to the number of positive Lyapunov exponents \citep{Ni2017}, which can bear a significant computational cost.
Third, the nonlinear dynamics of thermoacoustics can be non-hyperbolic, i.e. the covariant Lyapunov basis may become defective~\citep{Ruelle2009}, for certain design parameters~\citep{Huhn2020a}. 
This means that gradients cannot be guaranteed to exist for all thermoacoustic design parameters, which can hinder, and can even prevent, the optimization process via gradient-update. 
%
% 7. Gradient-free of this paper
%
In this paper, we develop a gradient-free optimization methodology to find the optimal design parameters that minimize the time-averaged acoustic energy.

% 8. Time average quantities 
In either gradient-based and gradient-free methods, the solution must be integrated sufficiently long in time (ideally, ad infinitum), such that the quantities of interest (gradient in gradient-based, and cost functional in gradient-free methods) have converged to within a desired precision\footnote{Both converge with $t^{-1/2}$, where $t$ is time.}. 
The generation of such a time series, however, can carry a high cost.
%
%
%
% 9. Machine learning of time series 
%
%
As an alternative to the integration of the governing equations, we propose the use of a data-driven technique to produce accurate predictions of the system's dynamics to generate the required long time series.
Such a task naturally falls within the category of supervised learning for time-dependent systems. 
In time-dependent problems, the order by which data is sorted (i.e. time) is of paramount importance. 
%If the order of the data is disregarded by a certain technique, then it does not make use of a crucial piece of information and its performance is bound to be worse than one that does take time into account.
Feed-forward neural networks are a classic architecture, which works well in regression problems \citep{Goodfellow2016}, but it does not naturally include recurrences to accurately learn the temporal correlations.  
To extend feed forward networks to sequential data, recurrent neural networks have an internal state, which is updated by taking into account both the current input and the previous state. 
Thus, the sequence by which data is fed affects the internal state, and therefore its output.
Within recurrent neural networks, three architectures are highlighted: 
(i) Long Short-Term Memory networks (LSTM) \citep{Hochreiter1997}, 
(ii) Gated Recurrent Unit networks (GRU) \citep{Cho2014}, and 
(iii) Echo State Networks (ESN) \citep{Jaeger2004,Lukosevicius2012}.
While the three have been successfully employed to learn and predict time-dependent problems, the ESN architecture offers an advantage, which is exploited in this paper. 
Because its output is a linear combination of the hidden state variables, its training reduces to a least-squares problem, which is more computationally robust than repeatedly calculating gradients, as in LSTM and GRU networks.
In chaotic learning, ESNs have been recently explored in multiple applications in chaotic systems, from time-accurate prediction \citep{Pathak2018,Doan2019}, to the reconstruction of hidden variables \citep{Lu2017,Doan2020,Racca2021b}, or the calculation of ergodic quantities \citep{Huhn2020c}.
%
%
% 10. ML in TA
% 
In particular, in \citet{Huhn2020c}, ESNs were employed in the prediction of the long-time average of a thermoacoustic dynamical system. 
Moreover, \citet{Hart2021} proved analytically that, under certain conditions, ESNs can approximate the invariant measure of a dynamical system, which is key to the calculation of accurate statistical quantities.
In this paper, we employ ESNs for the generation of sufficiently long time series, from which the time-averaged cost functional to be optimized is evaluated.
Specifically, we apply this framework to predict the time-averaged thermoacoustic energy. 

% 11. Objectives
The objective of this paper is three-fold. 
First, we propose a versatile gradient-free methodology to optimize time-averaged cost functionals. 
The methodology requires a minimal number of user-defined parameters, which makes it a minimally-intrusive tool. 
We apply the methodology to a chaotic thermoacoustic system. 
Second, we investigate the capability of ESNs of learning thermoacoustic solutions from small data. %~\lm{[Make sure to state later the size of the training data in Lyapunov times.}]. 
Both short- and long-time predictions are analyzed. 
Third, we minimize a chaotic thermoacoustic oscillation by finding the optimal set of design parameters. 

% 12. Structure 
The paper is structured as follows. 
Section~\ref{sec:problem_formulation} presents the general optimization problem with a focus on the thermoacoustic system.
Section~\ref{sec:methodology} introduces the proposed gradient-free optimization method, both in general and in particular case of this paper. The method combines the tools of Section~\ref{sec:bayesian_tools}, which describes Bayesian optimization; and Section~\ref{sec:echo_state_networks}, which presents both the traditional and hybrid echo state networks.
Section~\ref{sec:performance} investigates the short- and long-time predictions of the ESNs in learning thermoacoustic dynamics. 
Section~\ref{sec:chain_optimization} applies the framework of Section~\ref{sec:methodology} to the optimization of a chaotic thermoacoustic system.
A final discussion and conclusions end the paper.
\fh{We have also included a discussion of the potential cost benefit of the proposed optimization framework in Appendix~\ref{app:cost_analysis}.}

%=================
% SEC: Math models
%=================
\section{Problem formulation and physical models}
\label{sec:problem_formulation}
% 
%The gradient-free optimization methodology of this paper can be applied to a dynamical system governed by
We consider a nonlinear dynamical system 
\begin{equation}
    \label{eq:ode}
    \frac{d \bm q}{dt} = \bm F (\bm q, \bm s),
\end{equation}
where $\bm q \in \mathbb{R}^{N_d}$ is the state vector;
$\bm F: \mathbb{R}^{N_d} \rightarrow \mathbb{R}^{N_d}$ is a nonlinear operator;
 $\bm s \in \mathbb{R}^{N_p}$ is a vector of physical (or design) parameters; and 
${N_d}$ is the number of degrees of freedom of the system. 
Given an initial condition, $\bm q_0$, \cref{eq:ode} can be solved to obtain a solution $\bm q(t, \bm s)$.
We wish to optimize the time-average of a cost functional  
\begin{equation}
    \label{eq:avg_cost_func}
    \avgJ(\bm s) = \lim_{T\rightarrow\infty} \frac{1}{T} \int_0^T \J(\bm q(t, \bm s), \bm s) \; dt, 
\end{equation}
where $\J$ is, for example, an energy. Because we consider ergodic systems, $\langle\mathcal{J}\rangle$ does not depend on the initial condition or trajectory, but it depends only on the parameters, $\bm s$.
The goal is to find a set of parameters, $\bm s^+$, that minimize the time-averaged cost functional in Eq.~\ref{eq:avg_cost_func}. 
Mathematically, $\bm s^+$ is the solution of
\begin{align}
    \label{eq:optim_prob_1} &\underset{\bm s}{\min} \; \avgJ, \\
    \label{eq:optim_prob_2} &\bm G(\bm s) = 0, \\
    \label{eq:optim_prob_3} &\bm H(\bm s) \geq 0,
\end{align}
where $\bm G$ and $\bm H$ are equality and inequality constraints, respectively. 
The inequality constraints guarantee that the physical parameters are searched in a feasible region. 

\subsection{Thermoacoustic dynamical system}
We consider an acoustic resonator that consists of a tube and a heat source in it. 
We assume that the cut-off frequency of the acoustic resonator is sufficiently high such that only  longitudinal acoustics propagate. 
The mean flow is assumed to have a zero Mach number with a spatially averaged temperature. 
The equations that govern the acoustics are the linearized momentum and energy equations
\begin{align}
    &\dpartial{u}{t} + \dpartial{p}{x} = 0, \label{eq:rijke_continuous1} \\
	&\dpartial{p}{t} + \dpartial{u}{x} + \zeta p - \dot{q}\delta(x - x_f)= 0, \label{eq:rijke_continuous2}
\end{align}
where $u$, $p$, $\zeta$, $\dot q$, $\delta$ and $x_f$ are the acoustic velocity, pressure, damping, heat-release rate, Dirac delta and flame position, respectively, which are non-dimensionalized as in~\cite{Juniper2011,Huhn2020a}.
The axial coordinate is $x \in [0,1]$, which is non-dimensionalized by the tube length. 
The heat release rate is given by a modified King's law~\cite{King1914,Heckl1988,Polifke2001,Orchini2016}
\begin{equation}
    \dot{q}(t) = \beta [ \left(1+u(x_f, t-\tau)\right)^{1/2} - 1 ], \label{eq:kings_law}
\end{equation}
where $\beta$ and $\tau$ are the heat release intensity and time delay, respectively.
The time delay models the time that the heat release takes to be perturbed by an acoustic perturbation at the base of the heat source. 
The solutions are decomposed in $N_g$ acoustic eigenfunctions of the undamped acoustic system~\citep{Balasubramanian2008}, which is also known as Galerkin decomposition, 
\begin{align}
    &u(x,t) = \sum\nolimits_{j=1}^{N_g} \eta_j(t)\cos(j \pi x),  \\
    &p(x,t) = -\sum\nolimits_{j=1}^{N_g} \mu_j(t) \sin(j \pi x), \label{eq:galerkin_decomp}
\end{align}
which results in a system of $2N_g$ oscillators that are nonlinearly coupled by the heat source
\begin{align}
	% &\dot{\eta}_j - j \pi \mu_j = 0, \label{eq:rijke_discrete1} \\
	&\frac{d\eta_j}{dt} - j \pi \mu_j = 0, \label{eq:rijke_discrete1} \\
	% &\dot{\mu}_j + j \pi \eta_j + \zeta_j \mu_j + 2 \dot{q} \sin(j \pi x_f) = 0, \label{eq:rijke_discrete2}
	&\frac{d\mu_j}{dt} + j \pi \eta_j + \zeta_j \mu_j + 2 \dot{q} \sin(j \pi x_f) = 0, \label{eq:rijke_discrete2} 
\end{align}
where  
$\zeta_j = c_1 j + c_2 j^{1/2}$ is the modal damping, which damps out higher-frequency oscillations according to physical scaling~\cite{Landau1987}.
Despite its simplicity, the thermoacoustic model in \cref{eq:rijke_continuous1,eq:rijke_continuous2,eq:kings_law} qualitatively captures complex nonlinear dynamics and bifurcations, as shown in \cite{Balasubramanian2008,Subramanian2011,Juniper2011,Huhn2020a}. %\lm{[Do Balasubramanian and Juniper show the bifurcations? If not, leave them out.]}
%
%The time delay in \cref{eq:kings_law} makes this a time-delayed problem. 
Because we wish to use numerical integrators and echo state networks, which march from time step $n$ to $n+1$, it is convenient to transform the time-delayed problem~\cref{eq:kings_law} into an initial value problem.
To achieve this,
%To transform the time-delayed problem into an initial value problem~\lm{[Why do we do this here? Because the ESN is a data-driven time stepper from n to n+1. Cite Huhn.]},
we model the advection of a dummy variable $v$ with velocity $\tau^{-1}$ as~\cite{Huhn2020a}
\begin{align}
    \dpartial{v}{t} + \frac{1}{\tau}\dpartial{v}{X} &= 0,\quad 0 \le X\le 1, \label{eq:advection} \\
    v(X=0,t) &= u_f(t) \label{eq:advection_bc}. 
\end{align}
The time-delayed velocity is provided by the value of $v$ at the right boundary, i.e. $u_f(t-\tau) = v(X=1,t)$. Equation~\eqref{eq:advection} is discretized using $N_c + 1$ points with a Chebyshev spectral method~\citep{Trefethen2000}, which adds $N_c$ degrees of freedom.
Thus, these equations define a dynamical system with a state vector
%\lm{These equations define a dynamical system with state vector ...; ${\bm F}$ ...; and. ${\bm s}$ ... }
%
$
    \bm q = \left[\eta_1, \cdots, \eta_{N_g} , \mu_1, \cdots, \mu_{N_g}, v_1, \cdots, v_{N_c} \right]
$. 
This model, which is used as a proof of concept, qualitatively captures the key physics of nonlinear thermoacoustics~\citep{Huhn2020a}.
The cost functionals that we wish to obtain and minimize are the time averages of the acoustic energy and the Rayleigh index~\cite{Huhn2020b} 
\begin{align}
    \Eac(t) &= \int_0^1 \frac{1}{2}\left(u^2(t) + p^2(t)\right) \, dx = \frac{1}{4} \sum_{j=1}^{N_g} \left(\eta_j^2(t) + \mu_j^2(t)\right), \label{eq:Eac} \\
    I_\mathrm{Ra}(t) &= p_f(t) \dot q(t). %= \int_0^1 \zeta p^2(x,t) \, dx = \frac{1}{2} \sum_{j=1}^{N_g} \zeta_j \mu_j^2(t),
\end{align}
The first measures the total energy of the acoustic oscillations, while the second corresponds to the rate of input energy in the system, which is balanced out over time by the damping.
(As shown in~\cite{Huhn2020a}, the two time-averaged cost functionals are one-to-one related to each other, therefore, we focus on the optimization of the acoustic energy only.) 
%\lm{[Explain briefly the physical meaning of the Rayleigh index.]}
%
In this work, the following parameters are fixed: $x_f = 0.2$, $c_1 = 0.1$ and $c_2 = 0.06$~\cite{Subramanian2011}. Unless otherwise specified, we use 10 Galerkin modes (i.e. $N_g=10$) and 11 Chebyshev points (i.e. $N_c=10$), which provide a good compromise between accuracy and computational cost~\cite{Huhn2020a}. . We solve the equations numerically in time with a 3-stage Runge-Kutta solver~\cite{Kennedy2000}, with a time step of 0.01 time units.

\section{Gradient-free design optimization with echo state networks}
\label{sec:methodology}

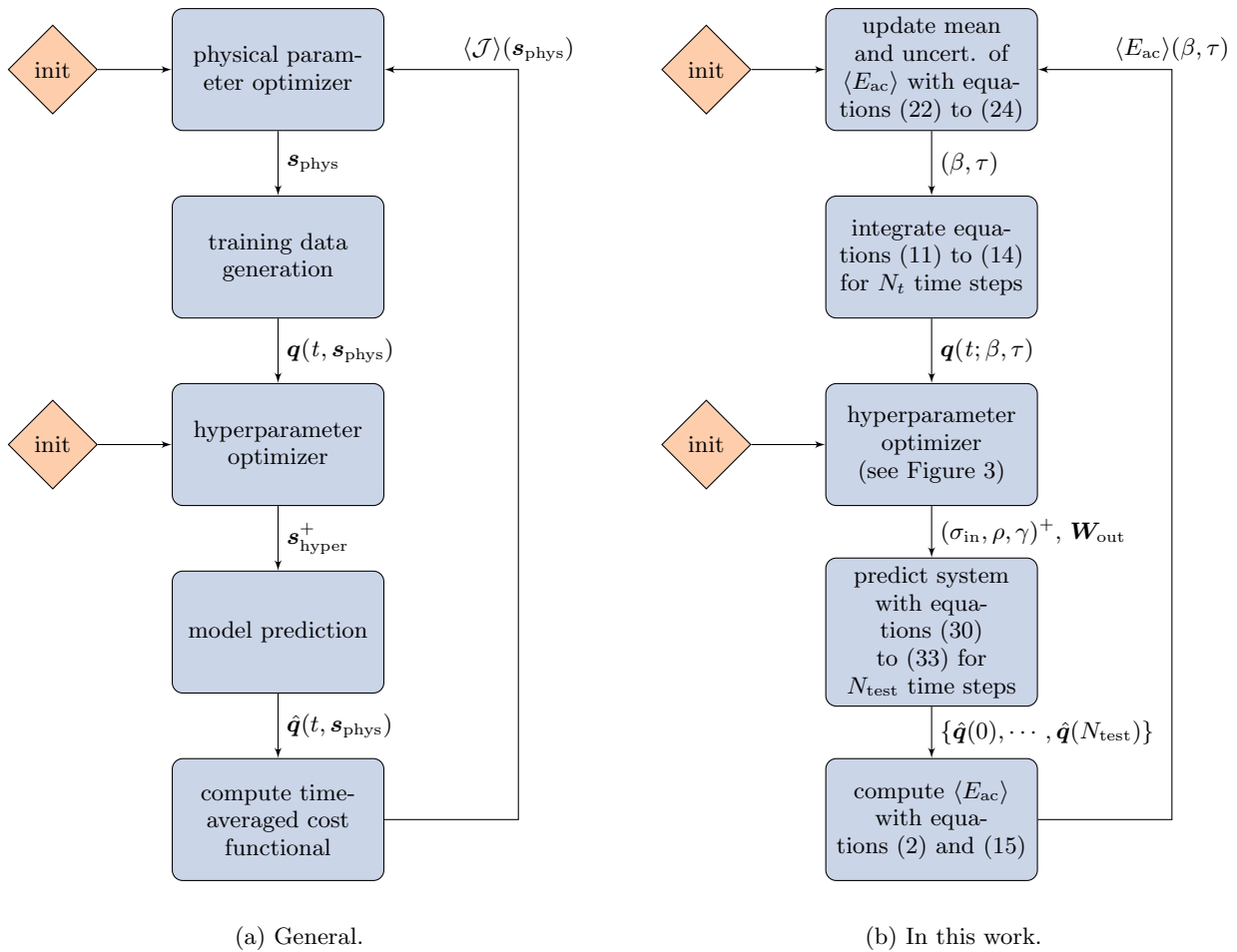
\begin{figure}[htb]
	\centering
	\begin{subfigure}{0.48\textwidth}
%	\tikzstyle{every node}=[font=\small]
	\definecolor{litegreen}{HTML}{B3E2CD}
	\definecolor{liteorange}{HTML}{FDCDAC}
	\definecolor{liteblue}{HTML}{CBD5E8}		
	% Define block styles
	\tikzstyle{decision} = [diamond, draw, fill=liteorange,%blue!20, 
	    text width=3em, text badly centered, node distance=3cm, inner sep=0pt]
	\tikzstyle{block} = [rectangle, draw, fill=liteblue,%blue!20, 
	    text width=8em, text centered, rounded corners, minimum height=5em]
	\tikzstyle{line} = [draw, -latex']
	\tikzstyle{cloud} = [ellipse, draw, fill=liteorange,%red!10,
	    text width=2em, text centered, node distance=3cm, minimum height=2.5em]
	
	\begin{tikzpicture}[node distance=2.5cm,auto]
	    % Place nodes
	    \node [block] (phys_optim) {physical parameter optimizer};
	    \node [block, below of=phys_optim] (generate_data) {training data generation};
	    \node [block, below of=generate_data] (hyper_optim) {hyperparameter optimizer};
	    \node [block, below of=hyper_optim] (prediction) {model prediction};
	    \node [block, below of=prediction] (computation) {compute time-averaged cost functional};
	    \node [decision, left of=phys_optim] (init_phys) {init};
	    \node [decision, left of=hyper_optim] (init_hyper) {init};
	    
	    % Draw edges
	    \path [line] (phys_optim) --  node {$\bm s_\mathrm{phys}$}(generate_data);
	    \path [line] (generate_data) -- node {$\bm{q}(\mathit{t},\bm{s}_\mathrm{phys})$} (hyper_optim);
	    \path [line] (hyper_optim) -- node {$\bm s_\mathrm{hyper}^+$} (prediction);
	    \path [line] (prediction) -- node {$\hat{\bm q}(t, \bm s_\mathrm{phys})$} (computation);
	    \path [line] (computation) -- ++(3.2,0) |- node [above] {$\langle \mathcal{J} \rangle(\bm s_\mathrm{phys})$} (phys_optim);
	    \path [line] (init_phys) -- (phys_optim);
	    \path [line] (init_hyper) -- (hyper_optim);
	\end{tikzpicture}
	\vspace{\baselineskip}
	\caption{General.}
    \label{fig:flowchart1}
    \end{subfigure}
    \begin{subfigure}{0.48\textwidth}
%	\tikzstyle{every node}=[font=\small]
	\definecolor{litegreen}{HTML}{B3E2CD}
	\definecolor{liteorange}{HTML}{FDCDAC}
	\definecolor{liteblue}{HTML}{CBD5E8}		
	% Define block styles
	\tikzstyle{decision} = [diamond, draw, fill=liteorange,%blue!20, 
	    text width=3em, text badly centered, node distance=3cm, inner sep=0pt]
	\tikzstyle{block} = [rectangle, draw, fill=liteblue,%blue!20, 
	    text width=8em, text centered, rounded corners, minimum height=5em]
	\tikzstyle{line} = [draw, -latex']
	\tikzstyle{cloud} = [ellipse, draw, fill=liteorange,%red!10,
	    text width=2em, text centered, node distance=3cm, minimum height=2.5em]
	
	\begin{tikzpicture}[node distance=2.5cm,auto]
	    % Place nodes
	    \node [block] (phys_optim) {update mean and uncert. of $\langle\Eac\rangle$ with \cref{eq:posterior2,eq:posterior3,eq:rbf}};
	    \node [block, below of=phys_optim] (generate_data) {integrate \cref{eq:rijke_discrete1,eq:rijke_discrete2,eq:advection,eq:advection_bc} for $N_t$ time steps};
	    \node [block, below of=generate_data] (hyper_optim) {hyperparameter optimizer (see \cref{fig:hyperparameter_selection})};
	    \node [block, below of=hyper_optim] (prediction) {predict system with \cref{eq:esn_feedback,eq:hesn_r,eq:hesn_y,eq:model_K} for $N_\mathrm{test}$ time steps};
	    \node [block, below of=prediction] (computation) {compute $\langle \Eac \rangle$ with \cref{eq:avg_cost_func,eq:Eac}};
	    \node [decision, left of=phys_optim] (init_phys) {init};
	    \node [decision, left of=hyper_optim] (init_hyper) {init};
	    
	    % Draw edges
	    \path [line] (phys_optim) --  node {$(\beta, \tau)$}(generate_data);
	    \path [line] (generate_data) -- node {$\bm{q}(\mathit{t}; \beta, \tau)$} (hyper_optim);
	    \path [line] (hyper_optim) -- node {$(\sigma_\subin, \rho, \gamma)^+$, $\Wout$} (prediction);
	    \path [line] (prediction) -- node {$\{\hat{\bm q}(0), \cdots, \hat{\bm q}(N_\mathrm{test})\}$} (computation);
	    \path [line] (computation) -- ++(3.2,0) |- node [above] {$\langle \Eac \rangle(\beta, \tau)$} (phys_optim);
	    \path [line] (init_phys) -- (phys_optim);
	    \path [line] (init_hyper) -- (hyper_optim);
	\end{tikzpicture}
	\vspace{\baselineskip}
	\caption{In this work.}
    \label{fig:flowchart2}
    \end{subfigure}
    \caption{Optimisation chain flowcharts. Human interaction is only required in the initialisation (\texttt{init} steps). The initialisation defines the search space, maximum number of evaluations, optimiser parameters, kernel functions, etc.}
    \label{fig:flowcharts}
\end{figure}

We introduce the proposed methodology to optimize a chaotic system with a non-intrusive approach, using echo state networks.
The flowchart in \cref{fig:flowchart1} illustrates the method. 
There are two optimizers, one for the physical parameters and another for the hyperparameters. 
The physical optimizer chooses the next point in the physical space to be evaluated. 
By integrating the ordinary differential equations that govern the thermoacoustic dynamics (ODEs), a short amount of data is generated, which is used to train the network. 
This mimics data from high-fidelity simulation or experiments, which is sparse and costly, the objective being to gain as much information as possible with a minimal number of samples.
Then, the hyperparameter optimizer selects the optimal hyperparameters. 
With the hyperparameters tuned, the data-driven model (echo state networks in this case) runs in predictive mode for a user-defined sufficiently long time to obtain the long-time average of the physical cost functional, which is returned to the physical parameter optimizer. % 
The only human intervention is at the start of the chain for the initialization of the optimizers.
After initialization (e.g. defining search space, maximum number of evaluations, etc.), the optimization chain runs on its own.
% %
% \begin{figure}[htbp!]
%     \centering
%     \includegraphics[width=0.5\textwidth]{img/flowchart.pdf}
%     \caption{Optimization chain flowchart. Human interaction is only required in the initialization (\texttt{init} steps), e.g.,   search space, maximum number of evaluations, optimizer parameters, kernel functions, etc. 
%     % 
%     }
%     \label{fig:flowchart}
% \end{figure}
%
\Cref{fig:flowchart2} depicts the chain of \cref{fig:flowchart1} in the present work.
In particular, in this work, the physical optimizer is a Bayesian optimizer using Gaussian Process Regression (GPR), with a Matérn 3/2 kernel.
Because of the nature of a Gaussian  Process~(\Sref{sec:gaussian_processes}), the dependent variable (in this case, the acoustic energy, $\Eac$) can be negative. 
Because a negative acoustic energy is unphysical, we apply the GPR to the logarithm of the acoustic energy, which means that values of the acoustic energy are modelled by a log-normal distribution. Hence, the estimates are positive and the standard deviation is additive in the exponent only.
Finally, because of interpretability, we use the Lowest Confidence Bound (see \Sref{sec:bayesian_optim}) as the acquisition function, with $\kappa=1.960$, which corresponds to a 95\% confidence interval. 
The ordinary differential equations are integrated using a 3-stage Runge-Kutta scheme~\citep{Kennedy2000}.
The hyperparameters are also selected via Bayesian optimization with GPR, but with an RBF kernel and GP-Hedge acquisition function.
The system is predicted with a hybrid echo state network model and the cost functional is the time-averaged acoustic energy, calculated with the prediction from the network.
All these concepts are introduced in the following two sections.

%We remark, however that, any gradient-free optimization algorithm and any data-driven model can be employed.

\section{Bayesian Optimization with Gaussian Process Regression}
\label{sec:bayesian_tools}
Gaussian Process Regression offers an estimate of both the mean and standard deviation of the cost functional.
This allows for a more informed choice to be made and for better control of the balance between exploration and exploitation (see \Sref{sec:bayesian_optim}).
Moreover, it is a global optimization method, which is advantageous when the cost functional is multimodal.
We summarize Gaussian Process Regression in \Sref{sec:gaussian_processes} and Bayesian Optimization in \Sref{sec:bayesian_optim}.

\subsection{Gaussian Process Regression}
\label{sec:gaussian_processes}

A Gaussian Process (GP) is a collection of random variables, any finite number of which have a joint Gaussian distribution \citep{Rasmussen}. 
Here, the random variables are the values of a function on its domain. In fact, the function is deterministic, but in the context of GPs, its (unknown) values are modelled as random variables.
A GP is specified by its mean function, $m(\bm x)$, usually set to 0, and the covariance function, often called kernel function, $k(\bm x, \bm x')$, which are defined as 
\begin{align}
    m(\bm x) &= \mathbb{E}\left[f(\bm x)\right], \\
    k(\bm x, \bm x') &= \mathbb{E}\left[\left(f(\bm x)-m(\bm x)\right)\left(f(\bm x')-m(\bm x')\right)\right],
\end{align}
where $f(\bm x)$ is the real process and $\mathbb{E}$ is the expectation. 
The Gaussian Process is written as
\begin{equation}
    f(\bm x) \sim \mathcal{G}\mathcal{P}\left(m(\bm x), k(\bm x, \bm x')\right), 
\end{equation}
where $\{(\bm x_i, f_i) | i = 1, \dots, n\}$ is a collection of $n$ data points, from which we construct the output vector $\bm f$ and the matrix $\bm X$, whose columns are the vectors $\bm x_i$. 
Similarly, we can define $\bm f_*$ and $\bm X_*$ for the $n_*$ test inputs, i.e. the inputs that we wish to predict. 
According to the definition of a GP, prior to any observations, the joint distribution of $\bm f$ and $\bm f_*$ is a Gaussian distribution,
\begin{equation}
    \begin{bmatrix}
        \bm f \\ \bm f_*
    \end{bmatrix} \sim
    \mathcal{N}\left(\bm 0,
    \begin{bmatrix}
        \bm K(\bm X, \bm X) & \bm K(\bm X, \bm X_*) \\
        \bm K(\bm X_*, \bm X) & \bm K(\bm X_*, \bm X_*)
    \end{bmatrix}
    \right),
\end{equation}
where, for any $\bm X_1$ and $\bm X_2$, $\bm K(\bm X_1, \bm X_2)$ is an $n_1 \, \times \, n_2$ matrix of covariances evaluated for all pairs of columns (each column being one point) of $\bm X_1$ and $\bm X_2$.
To include the information from the observations $f_i$, this distribution is conditioned on the observations \citep{Rasmussen} 
\begin{align}
    \bm f_*|\bm X_*,\bm X, \bm f  &\sim \mathcal{N} ( \bm \mu_*, \bm \Sigma_* ), \label{eq:posterior1} \\
    \bm \mu_* &= \bm K(\bm X_*, \bm X) \bm K(\bm X, \bm X)^{-1} \bm f, \label{eq:posterior2} \\
    \bm \Sigma_* &= \bm K(\bm X_*, \bm X_*) - \bm K(\bm X_*, \bm X)\bm K(\bm X, \bm X)^{-1}\bm K(\bm X, \bm X_*). \label{eq:posterior3}
\end{align}
If the observations are noisy, with variance $\sigma_n^2$, then $\bm K$ is replaced by $\bm K + \sigma_n^2 I$, where $I$ is the identity matrix.

In this work, we use two kernel functions, the Radial Basis Function, %\lm{[Justify why we use this]}
\begin{equation}
    k(\bm x, \bm x') = \exp\left(-\frac{||\bm x - \bm x'||^2}{2 l^2}\right),
    \label{eq:rbf}
\end{equation}
and the Matérn 3/2 Kernel Function,%  \lm{[Justify why we use this]}
\begin{equation}
    k(\bm x, \bm x') =  \left(1 + \frac{\sqrt{3}}{l} ||\bm x - \bm x'||\right) \exp \left(-\frac{\sqrt{3}}{l} ||\bm x - \bm x'|| \right),
    \label{eq:matern}
\end{equation}
where $|| \cdot ||$ is the Euclidean distance and $l$ are tunable length scales, which control the smoothness of the function being regressed. 
A large $l$ means that the covariance will be high even for relatively distant points, $\bm x$ and $\bm x'$, resulting in a smoother function than that for smaller $l$.
Because we expect the acoustic energy
to be smoother in the physical space~\citep{Huhn2020a} than the mean squared error in the hyperparameter space~\citep{Racca2021}, we use the RBF kernel for the former and the Matérn kernel for the latter.
%The reason why different kernels are used is the expectation of the space of the functions. We expect the acoustic energy to be smoother in the physical space~\citep{Huhn2020a} than the MSE in the hyperparameter space~\citep{Racca2021}.
%Thus, we use the RBF kernel, which is smoother than the Matérn 3/2 kernel, in the physical parameter optimizer, and the Matérn kernel in the hyperparameter optimizer.

\subsection{Bayesian Optimization}
\label{sec:bayesian_optim}

Bayesian optimization is used in this work to select the hyperparameters, and to optimize the physical parameters such that the acoustic energy of the system is minimal.

Bayesian optimization consists of a loop of three steps:
\begin{enumerate}
    \item Observe a point from the optimization domain;
    \item Update the mean, $\mu_*$, and uncertainty, $\Sigma_*$;
    \item Select the next point to observe by finding the minimum of the acquisition function.
\end{enumerate}
The first step simply evaluates the function, $f$, at the given point $\bm x$.
The second step calculates the mean and variance of the distribution of \cref{eq:posterior2,eq:posterior3}.
Finally, in the third step, the new point to observe corresponds to the optimum of an acquisition function.

An acquisition function takes into account both the mean and uncertainty and, for any point $\bm x$, outputs a value that relates to either or a combination of the two. This provides the probability, or amount, by which $\bm x$ can improve the current optimum. 
There are four common acquisition functions: 
Probability of Improvement (PI), Expected Improvement (EI), Lowest Confidence Bound (LCB), and GP-Hedge~\citep{Hoffman2011}.
The first, PI,  computes the probability that a candidate $\bm x$ can improve with respect to the current optimum, $\mathrm{prob}(f(\bm x) < f^+)$.
The second, EI, is similar to PI, but weighs the probability by the potential gain, i.e. it is the expected value of the improvement, $\mathbb{E}[\max(f^+-f(\bm x), 0)]$.
Finally, LCB is based on intervals of confidence centered around the mean, $[\mu - \kappa \sigma, \mu + \kappa \sigma]$.
For a given value of $\kappa$, the interval will cover a certain percentage of the outcomes. 
For example, with $\kappa \approx 1.960$, 95\% of the outcomes will be contained in the interval, i.e. a 95\% confidence interval\footnote{For an $\alpha$\% confidence interval, one finds $\kappa$ such that $\Phi(\mu + \kappa \sigma) - \Phi(\mu - \kappa \sigma) = \alpha$, where $\Phi$ is the cumulative distribution function of the Gaussian distribution $\mathcal{N}(\mu, \sigma^2)$.}.
When the LCB acquisition function is used, one seeks to find $\bm x$ for which the lower bound of the confidence interval is a minimum, i.e. the $\bm x$ with the smallest lower confidence bound. 
This means that $\kappa$ controls the balance between {\it exploration} (exploring unobserved regions of the optimization space) and {\it exploitation} (improving an existing observation by searching close to it), with exploration being preferred when $\kappa$ is large ($\kappa$ multiplies the uncertainty $\sigma$) and exploitation when $\kappa$ is small.
Finally, GP-Hedge \citep{Hoffman2011} overcomes the difficulty of knowing which acquisition function will perform best by taking the previous three acquisition functions\footnote{GP-Hedge can be applied to any combination of acquisition functions, not only PI, EI and LCB.} and probabilistically picking one of the three suggestions to sample next. 
For each acquisition function, the better the means, $\mu$, of the points suggested in the past, the larger the probability of being chosen.
In this paper, we use the LCB because, although it does not necessarily offer the best performance among the four acquisition functions, it is simple to compute and, more importantly, it is simple to physically interpret.
%
%

% =====================
%SEC: ESN
% =====================
\section{Echo State Networks}
\label{sec:echo_state_networks}
%\lm{[These tools need to be introduced in the introduction to a larger extent.]}
Echo state networks (ESN)~\cite{Jaeger2004,Lukosevicius2012} are recurrent neural networks, which are composed of a set of nodes that constitute the reservoir. 
The ESN receives an input signal, $\bm l(n) \in \R^{N_l}$, and produces an output signal, $\hat{\bm y}(n) \in \R^{N_y}$, where $n$ is the discrete time variable, i.e. $t=n \, \Delta t$. 
Usually $N_l = N_y = N_d$, where $N_d$ is the dimension of the system being predicted, such that the network can evolve on its own.
The state of the reservoir is a vector, $\bm r$, of the states of all units, $r_j, \; j \in \{1,\dots,N_r\}$. 
The reservoir state evolves according to the nonlinear law
\begin{equation}
    \label{eq:reservoir_eq}
    \bm r(n) = \tanh(\Win [\bm l(n); b_\subin] + \W \bm r(n-1)),
\end{equation}
where $\Win$ is the input matrix (i.e. $\Win^{i,j}$ is the weight from the $j$-th component of the input to the $i$-th node) 
and $b_\subin$ is the input bias, with the semicolon denoting row concatenation.
Similarly, in the recurrency matrix $\W$, the component $\W^{i,j}$ is the weight from the $j$-th node to the $i$-th node. 
Therefore, $\Win$ and $\W$ are $N_r \times N_d$ and $N_r \times N_r$ matrices. 
The hyperbolic tangent in \cref{eq:reservoir_eq} is applied entry-wise.
Finally, the output is calculated by linear combination of the states of the reservoir units,
\begin{equation}
    \label{eq:esn_output}
    \hat{\bm y}(n) = \Wout [\bm r(n); b_\subout],
\end{equation}
where $\Wout$ is the output matrix, of size $N_d \times N_r$, and $b_\subout$ is the (scalar) output bias.

The network is trained to produce an output $\hat{\bm y}$ that matches the target $\bm y$ by minimizing %a metric. Here, the metric is
the mean square error (MSE)~\citep{Jaeger2004,Lukosevicius2012} 
\begin{equation}
    \label{eq:MSE}
    \mathrm{MSE} = \frac{1}{N_t} \sum_{n=1}^{N_t} \frac{||\hat{\bm y}(n) - \bm y(n)||^2}{N_d},
\end{equation}
where $N_t$ is the number of (discrete) time steps, 
and the norm is Euclidean in $\R^d$.
In ESNs, both $\Win$ and $\W$ are generated once and fixed.
In this work, each reservoir node is connected to one input, which results in every row of $\Win$  having only one non-zero entry.
The weight of the connections is sampled from a uniform distribution in the range $[-\sigma_\subin, \sigma_\subin]$, where $\sigma_\subin$ is a scaling parameter. Hence, $\Win$ can be generated by sampling uniformly from the range $[-1,1]$ and scaling by $\sigma_\subin$ directly in \cref{eq:reservoir_eq}.
Similarly, $\W$ is generated by sampling from the uniform distribution in the range $[-1,1]$, with each node being on average connected to $(1-\mathrm{sp}) N_r$ other nodes, where $\mathrm{sp}$ is the desired sparseness.
The matrix is then scaled to have a desired spectral radius, $\rho$, which is typically smaller than unity to satisfy the echo state property~\citep{Jaeger2004,Lukosevicius2012}.
A network that satisfies the echo state property ``forgets'' an old input after a certain time, which means that, even if starting from two different states, a network with the echo state property will converge to the same trajectory after a certain time (provided it is fed by the same input).

\begin{figure}[tb]
    \centering
    \includegraphics[width=0.7\textwidth]{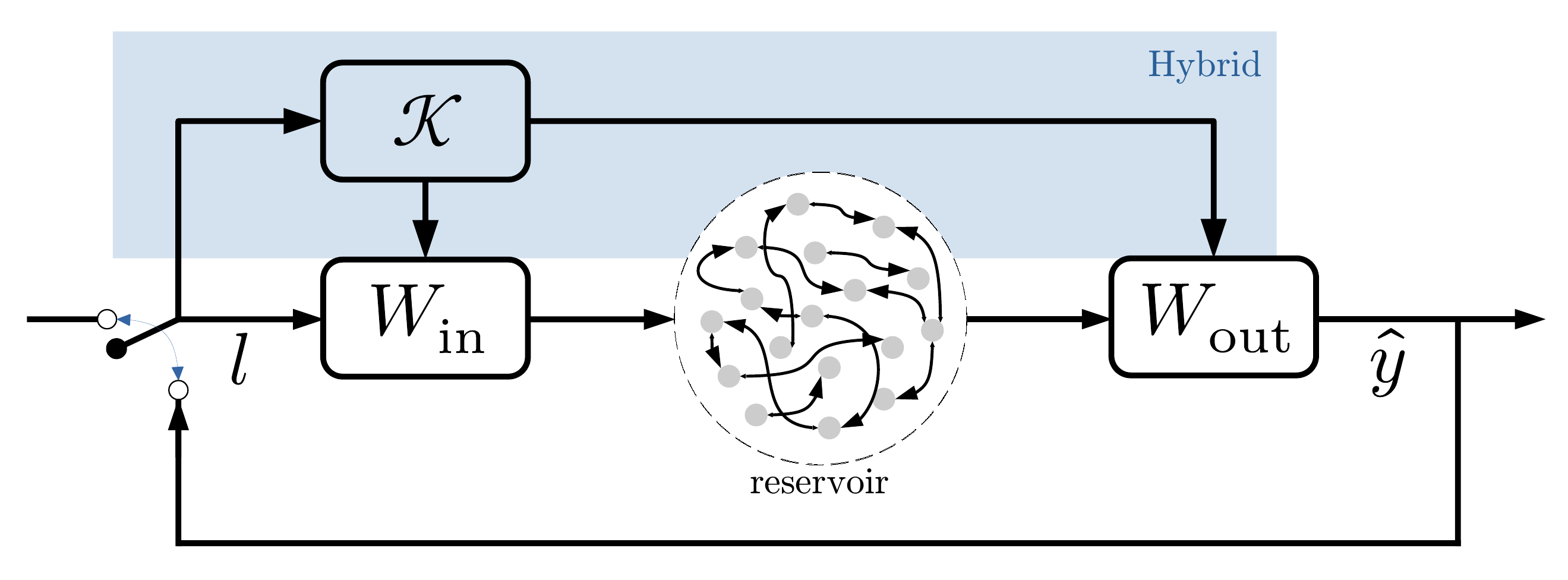}
    \caption{Schematic of the conventional (without highlighted region) and hybrid echo state networks. The hybrid echo state network corresponds to a conventional ESN with an additional knowledge-based model ($\K$ box) that feeds the reservoir and the output via the augmented $\Win$ and $\Wout$ matrices. In training, the switch is horizontal, whereas in prediction, it is vertical. %\lm{[Move this up to the ESN section. See comment after eq 10.]}
    }
    \label{fig:hESN_schematic}
\end{figure}

Because $\Win$ and $\W$ are fixed, only the output weights (i.e. the entries of $\Wout$) are trained to solve the minimization problem \cref{eq:MSE} 
with ridge regression 
\begin{equation}
    \label{eq:Wout}
    \Wout = (\bm R^T \bm R + \gamma \bm I)^{-1} \bm R^T \bm Y,
\end{equation}
where $\gamma$ is  the user-defined Tikhonov factor, which regularises the training. 
The $\bm R$ and $\bm Y$ matrices are obtained by row-concatenating the reservoir states and output targets, i.e. the $n$-th row corresponds to the discrete time $n$. 
During training mode, the network is operated in open-loop, whereas in prediction mode, the output of the network is fed to its input (closed-loop), i.e.
\begin{equation}
    \bm l(n+1) = \hat{\bm y}(n),
    \label{eq:esn_feedback}
\end{equation}
for the network to evolve autonomously.
This corresponds to the schematic of \cref{fig:hESN_schematic} with the blue highlighted region removed.
% \lm{[Refer to fig. 1 for a schamtic (K=0). Add in the caption the distinction between ESN (K=0) and hESN (K not 0).]}

% H-ESN 
\subsection{Hybrid Echo State Network}
\label{sec:hybrid_esn}

The hybrid echo state network (hESN) is a variant of the conventional ESN~\cite{Pathak2018}. 
In the hESN, the capability of the conventional ESN is complemented by (possibly imperfect) physical knowledge from a dynamical system, which may be a reduced-order model. 
The combination of data and model knowledge achieves higher accuracy, not only in the short time prediction \cite{Pathak2018,Racca2021}, but also in the long time \cite{Huhn2020c}. 
\Cref{fig:hESN_schematic} shows the architecture of an hESN. 
The network's input is fed to the reservoir, as in the conventional ESN, and to the physical knowledge based system (marked $\K$). 
The output of the physical system, in turn, is passed to both the reservoir, via $\Win$; and the output, via $\Wout$.
Mathematically, \cref{eq:reservoir_eq} and \cref{eq:esn_output} are augmented by $\K$
\begin{align}
    \bm r(n) &= \tanh(\Win [\bm l(n); b_\subin; \K(n)] + \W \bm r(n-1)), \label{eq:hesn_r}\\
    \hat{\bm y}(n) &= \Wout [\bm r(n); b_\subout; \K(n)] \label{eq:hesn_y},
\end{align}
where the dependence on $n$ is implicit via the input, $\bm l(n)$
\begin{equation}
    \K(n) = \K(\bm l(n)).
    \label{eq:model_K}
\end{equation}
Although the hESN can perform better than a conventional ESN of equal size, as shown in \Sref{sec:performance}, it can result in an unstable behavior. 
In prediction mode, the feedback of the output of $\K$, via $\Wout$, into its own input can create a self-sustaining amplification that diverges to infinity (see Appendix~\ref{app:divergence}). 
In such cases, validation and test errors are undefined, violating regularity assumptions (e.g. continuity in the hyperparameter space), which are essential to many optimization algorithms.
We propose ways of overcoming this issue.
The first is error saturation. 
If the prediction error becomes greater than a threshold, the error is set to the threshold.
The second is saturation, where a saturation function is applied to the output of the physical model itself, e.g. $\K \rightarrow \tanh(\K)$, with the $\tanh$ taken entry-wise (as in the conventional reservoir update equation).
This can be seen as effectively increasing the reservoir by a number of units equal to the dimension of $\K$, where each of these units is connected to one entry of $\K$ only and without repetition. 
The drawback is that, due to the saturation, the sensitivity to changes in the output of $\K$ is reduced, which can impact the performance.
The third is to eliminate the connection between $\K$ and $\Wout$, with the output of $\K$ feeding the reservoir only, effectively preventing unbounded growth.
We tested the three suggested methods (result not shown). We found that the first option performed best for the case under investigation, which is why it is adopted in the remainder of the paper.

\subsection{Hyperparameter selection}
\label{sec:hyperparameter_selection}

The traditional technique for selecting hyperparameters is manual selection, which is dependent on prior (human) knowledge and experience. 
However, that does not suit a non-intrusive approach, which is central to the objective of this work, as explained in~\Sref{sec:chain_optimization}.
The simplest non-intrusive technique is grid search, but it can carry high computational cost~\citep{Yperman2016,Maat2019,Racca2021}.  
Furthermore, the discretization of the hyperparameter space is a delicate matter because, if it is too coarse, the optimum can be missed; whereas, if it is too fine, the computational cost becomes prohibitive.
%Thus, more efficient techniques, which make use of reasonable assumptions about the quantity to minimize, as well as taking into account already known values of that same quantity, should be explored for this task.
Bayesian optimization with Gaussian Process Regression has been documented to achieve good performance in hyperparameter tuning of echo state networks~\cite{Racca2021}.
For example, \citet{Maat2019} found that this technique systematically achieves similar, or lower, values of test error compared to grid search, but with fewer evaluations.
In an in-depth examination of training techniques \cite{Racca2021} (e.g. single-shot, cross validation, etc.), it was found that grid search and Bayesian optimization have similar values of validation error, with Bayesian optimization being more robust and efficient. 
In this paper, the hyperparameters are selected by minimizing the validation mean square error (validation MSE), using Bayesian optimization with Gaussian Process Regression (GPR). 
We use the implementation in the \texttt{scikit-optimize} library.
The GPR uses a 3/2 Matérn Kernel~\citep{Rasmussen} (see \Sref{sec:gaussian_processes}) and the acquisition function is the GP-Hedge (see \Sref{sec:bayesian_optim}). 
The initial seed points are generated using a Latin Hypercube Sampling method. 
To make it more amenable to optimisation, we smooth the cost functional with two modifications.
First, because the values of the validation MSE cover multiple orders of magnitude (e.g. $10^{-6}$ to $10^3$), we minimize the logarithm of the MSE.
Second, we cap the error when it is larger than the threshold of $10^3$, as explained in \Sref{sec:hybrid_esn}.
The saturation smooths the MSE at these points.

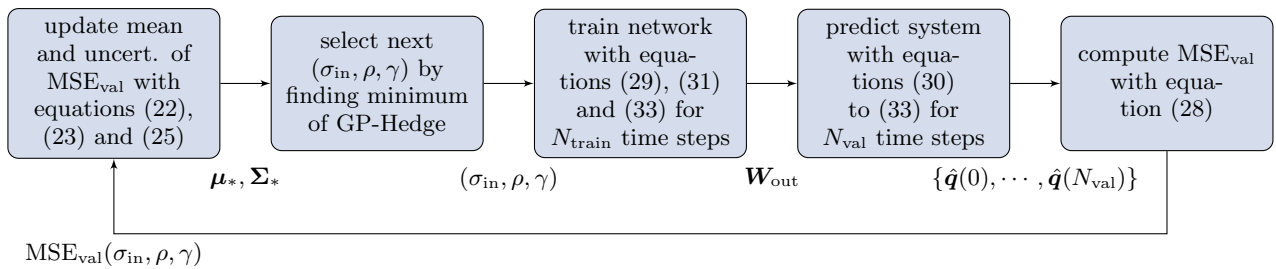
\begin{figure}[htb]
    \centering
%	\tikzstyle{every node}=[font=\small]
	\definecolor{litegreen}{HTML}{B3E2CD}
	\definecolor{liteorange}{HTML}{FDCDAC}
	\definecolor{liteblue}{HTML}{CBD5E8}		
	% Define block styles
	\tikzstyle{decision} = [diamond, draw, fill=liteorange,%blue!20, 
	    text width=3em, text badly centered, node distance=3cm, inner sep=0pt]
	\tikzstyle{block} = [rectangle, draw, fill=liteblue!80,%blue!20, 
	    text width=8em, text centered, rounded corners, minimum height=5.5em]
	\tikzstyle{line} = [draw, -latex']
	\tikzstyle{cloud} = [ellipse, draw, fill=liteorange,%red!10,
	    text width=2em, text centered, node distance=3cm, minimum height=2.5em]
	
	\begin{tikzpicture}[node distance=3.5cm,auto]
	    % Place nodes
	    \node [block] (optim) {update mean and uncert. of $\mathrm{MSE}_\mathrm{val}$ with \cref{eq:posterior2,eq:posterior3,eq:matern}};
	    \node [block, right of=optim] (acq) {select next $(\sigma_\subin, \rho, \gamma)$ by finding minimum of GP-Hedge};
	    \node [block, right of=acq] (train) {train network with \cref{eq:Wout,eq:hesn_r,eq:model_K} for $N_\mathrm{train}$ time steps};
	    \node [block, right of=train] (predict) {predict system with \cref{eq:esn_feedback,eq:hesn_r,eq:hesn_y,eq:model_K} for $N_\mathrm{val}$ time steps };
	    \node [block, right of=predict] (compute) {compute $\mathrm{MSE}_\mathrm{val}$ with \cref{eq:MSE}};
	   % \node [block, below of=hyper_optim] (prediction) {closed-loop prediction};
	   % \node [decision, left of=phys_optim] (init_phys) {init};
	   % \node [decision, left of=hyper_optim] (init_hyper) {init};
	   % % Draw edges
	    \path [line] (optim) --  node[below=1cm] {$\bm \mu_*, \bm \Sigma_*$}(acq);
	    \path [line] (acq) -- node[below=1cm] {$(\sigma_\subin, \rho, \gamma)$} (train);
	    \path [line] (train) -- node[below=1cm] {$\Wout$} (predict);
	    \path [line] (predict) -- node[below=1cm] {$\{\hat{\bm q}(0), \cdots, \hat{\bm q}(N_\mathrm{val})\}$} (compute);
	    \path [line] (compute) -- ++(0,-2) -| node [below] {$\mathrm{MSE}_\mathrm{val}(\sigma_\subin, \rho, \gamma)$} (optim);
	   % \path [line] (init_phys) -- (phys_optim);
	   % \path [line] (init_hyper) -- (hyper_optim);
	\end{tikzpicture}
    \caption{Hyperparameter selection with Bayesian optimization.}
    \label{fig:hyperparameter_selection}
\end{figure}

The minimization runs until the MSE is below a target threshold of $3 \times 10^{-2}$, which was chosen by trial and error; or the maximum number of calls, 20, has been reached. This value for the maximum number of calls is sufficiently large for the hyperparameter space to be explored, but not too large for the computation to become exceedingly expensive. 
%Other more sophisticated stopping criteria could have been employed, such as stopping when an arbitrary number of the last calls has not improved the result.
% 
We optimally tune $\rho$, $\sigma_\subin$, which are two hyperparameters that markedly affect the training~\citep{Racca2021}.
The hyperparameter space is log-uniform, i.e. the optimization tunes the exponents of the hyperparameters,% 
\begin{align}
    \log_{10}(\sigma_\subin) &\in [-2, 2], \\
    \log_{10}(\rho) &\in [-3, 0].
\end{align}
A log-uniform allows a more efficient exploration of different scales than a linear space. In particular, $\rho$ is related to the time scale of the dynamics, which can be of different orders of magnitude for different systems or attractors.
In the optimization of \Sref{sec:chain_optimization}, because the attractor varies, we also include the Tikhonov factor, $\gamma$, as a hyperparameter with $\log_{10}(\gamma) \in [-11, -4]$.
The process of hyperparameter selection via Bayesian optimization is schematized in \cref{fig:hyperparameter_selection}.

\subsection{Data normalisation}

Data normalization is crucial to obtaining good performance~\citep{Lukosevicius2012}. Because different components of the data vector can have vastly different ranges, a single input scaling factor, $\sigma_\subin$, can be insufficient.
If the same scaling is applied to variables of different orders of magnitude, the $\tanh$ might ``ignore'' one of the variables because of the saturation of values away from 0. In that case, the information from that variable would be lost.

While various normalizations exist, here we choose the ``min-max'' normalization, which divides the data variable by the difference between its maximum and minimum in the time period.
For an unnormalized variable, $\breve{l}$, whose time series is $\{\breve{l}(0), \breve{l}(1), \dots\}$, the normalised variable, $l$, is given by
\begin{equation}
	l(n) = \frac{\breve{l}(j)}{\max\limits_j\left(\breve{l}(j)\right) - \min\limits_j\left(\breve{l}(j)\right)}, \quad n = 0, 1, \dots.
\end{equation}
This normalisation forcibly makes $l \in [-1, 1]$, which means that all the variables have the same range.
%Therefore, it brings all variables to a similar range and the use of single scaling factor becomes justified.
With all the (normalized) variables in the same range, the use of the single scaling factor $\sigma_\subin$ is justified.

\section{Results}

The training, validation and test lengths are $N_\mathrm{train} = 5000$, $N_\mathrm{validation} = 2000$ and $N_\mathrm{test} = 10000$, respectively.
For the chaotic attractor of \Sref{sec:short-time_performance}, this corresponds to approximately 6, 2.4 and 12 Lyapunov times\footnote{The leading Lyapunov exponent is approximately 0.12~\cite{Huhn2020c}, which means that the Lyapunov time is approximately $0.12^{-1}$.}.
%\lm{[How do they compare with Lyapunov times?]}
To initialize the network, the reservoir state is initialized to $0$ and the first 100 iterations are discarded. This ensures that the reservoir of network is not affected by the initialization.
%\lm{[Explain why we need to wash out.]}
The physical model of the hESN is the same dynamical system that generates the data (\cref{eq:rijke_discrete1,eq:rijke_discrete2}), but with one Galerkin mode only instead of 10 (i.e. $N_g=1$).
This mimics a situation in which data is available from an experiment for which a simple physical model exists. 
Alternatively, data can come from a high-fidelity simulation, while $\K$ is a reduced-order model obtained from first principles and approximations.

The reservoir is composed of 400 and 100 nodes in the conventional and hybrid architectures, respectively.  % \lm{[We need a convergence test on this.]}
Although the reservoir size is usually chosen by heuristics, such as choosing the largest one can afford \citep{Lukosevicius2012}, or by human experience, here, for completeness, we show that these values are optimal for their respective architectures.
To analyse the quality of a prediction, we use the Kullback-Leibler divergence \citep{Kullback1951}
\begin{equation}
    D_\mathrm{KL}(P || Q) = \int p(\bm x) \log\left(\frac{p(\bm x)}{q(\bm x)}\right) dV,
    \label{eq:D_KL}
\end{equation}
where $P$ and $Q$ are continuous distributions and $p$ and $q$ are the respective probability density functions. The integral is taken in the phase space of the system.   In the Kullback-Leibler divergence, $P$ refers to a ``truth'', against which a ``model'' $Q$ is being compared, which is suitable for the present situation, in which the truth is the data generated by the ODEs and the models are the echo state networks.
If $Q$ perfectly matches $P$, i.e. $p(\bm x) = q(\bm x) \; \forall \, \bm x$, then $D_\mathrm{KL} = 0$, indicating that the larger $D_\mathrm{KL}$, the worse the match. The numerical calculation of~\cref{eq:D_KL} is performed with the empirical distributions, i.e. via the histograms of $P$ and $Q$.
\begin{figure}[htbp!]
    \centering
    \includegraphics[width=\textwidth]{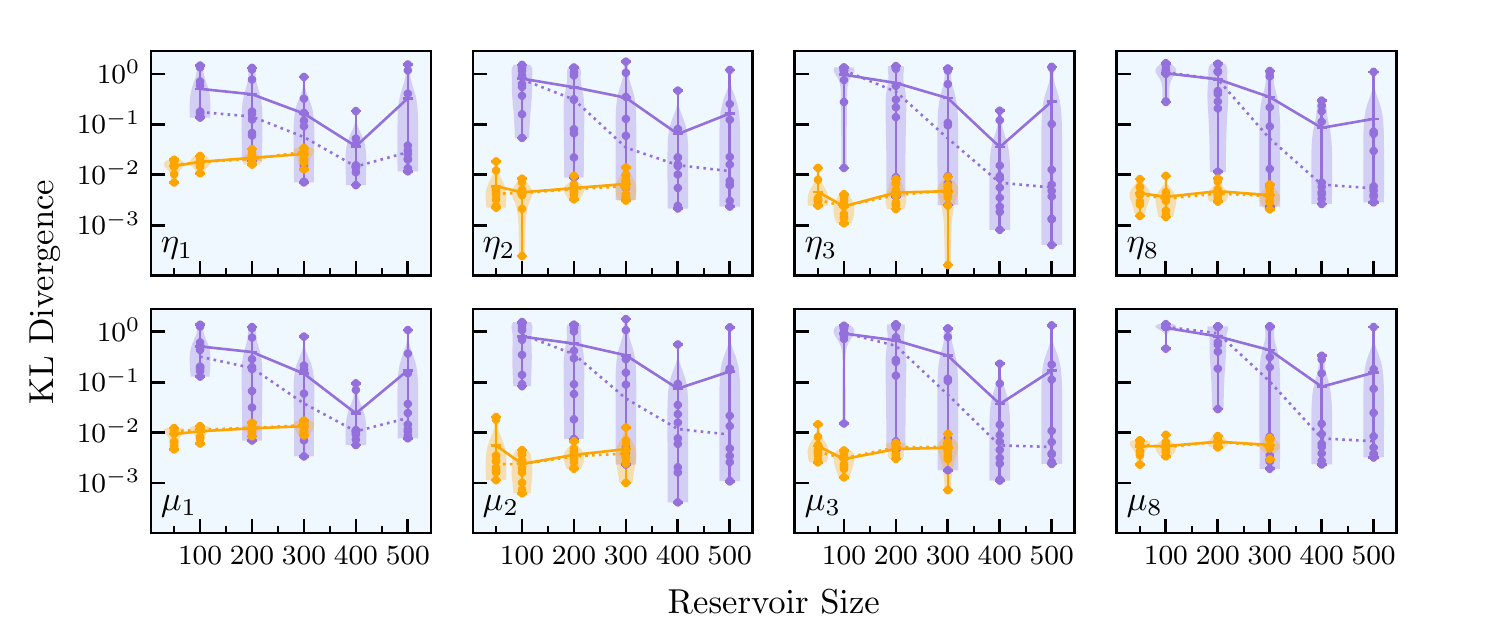}
    \caption{Kullback-Leibler divergence, $D_{KL}$, versus reservoir size.  The first and second rows correspond to the velocity and pressure modes, $\eta_j$ and $\mu_j$. The shaded regions correspond to the distributions of $D_\mathrm{KL}$ arising from the reservoir realizations. Each dot is a reservoir realization. The solid and dashed lines correspond to the mean and median.}
    \label{fig:kl_convergence}
\end{figure}

\Cref{fig:kl_convergence} shows the variation of $D_\mathrm{KL}$ with respect to the reservoir size for both the conventional and hybrid echo state network architectures. For each reservoir size, an ensemble of 10 network realizations is run, which allows the estimation of a distribution with its mean and spread. This is shown for the Galerkin modes 1, 2, 3 and 8, with the first three corresponding to the three most energetic modes and the last being representative of higher order modes.
The hESN performs better on average than the ESN. In fact, there are only a few realizations of the ESN that outperform realizations of the hESN. This is not unexpected and is further explored in the next section. Furthermore, compared to the hESN, the ESN exhibits much larger variability within realizations of the same size.
Analyzing separately, the hESN exhibits low variability in network realizations. The results also indicate that reservoir size has little, and possibly detrimental, impact on performance, with the optimal size being 100. 
%This might seem counter-intuitive, especially when taking common heuristics such as using the largest reservoir one can afford \citep{Lukosevicius2012}. 
This is due to a combination of: i) the physical model, which offers estimates of a good quality leaving reduced work to the network; ii) too many reservoir nodes, thus training parameters, for the small amount of data used in the training.
On the one hand, at very low numbers of nodes, the network will have high error because it does not have a sufficient number of parameters to train. On the other hand, at very large number of nodes, there are too many parameters and overfitting becomes a problem. Therefore, a U-shaped curve can be expected. This shape is barely visible for hESN because there is only one point, $50$, to the left of the optimum.
Notwithstanding, this effect is more visible in the ESN curves, where there is an improvement as the reservoir size increases. Although both mean and median decrease up to $400$, the mean increases from $400$ to $500$, while the median flattens or slightly decreases. This is explained by the large variability of the reservoirs of size $500$.
Therefore, given the similar median and comparable performances between ESN realizations of sizes $400$ and $500$, we select $400$ nodes to keep the computational cost minimal. 
The realizations chosen for the following section are those closest to the median of selected sizes.

\subsection{Short- and long-time predictions}
\label{sec:performance}

In this section, we compare the predictive capabilities of both the conventional % 
(ESN) and hybrid echo state networks (hESN).
We fix the physical parameters $\beta=7.0$ and $\tau=0.2$, which correspond to a chaotic solution~\cite{Huhn2020c}. 
The physical system (\cref{eq:rijke_discrete1,eq:rijke_discrete2}) will be referred to as the ``Truth''. % 
Information about each network, including the optimal hyperparameters, is given in \cref{tbl:optimal_hyperparameters}.

On the one hand, in short-time prediction, the objective is to time-accurately reproduce the dynamics of the system, i.e. starting from some initial condition, the objective is for the difference between the prediction and the true signals to be minimal for the largest possible time. This task is covered in \Sref{sec:short-time_performance}.
On the other hand, in long-time prediction, the objective is to accurately reproduce the ergodic properties (statistics) of the system, i.e. the objective is for the difference between the true attractor (the stationary measure) and the attractor of the echo state networks to be minimal.
Good performance in either task does not necessarily imply good performance in the other, as can be seen in \Sref{sec:long-time_performance} (good long-, but poor short-time performance with the conventional ESN) and Appendix~\ref{app:symmetric_lorenz} (good short-, but poor long-time performance).

\begin{table}[b]
    \centering
    \setlength{\tabcolsep}{8pt}
    \begin{tabular}{ l|cccc|cc }
        &  $N_r$ & $\bm b_\subin$ & $\bm b_\subout$ & $\mathrm{sp}$ & $\rho$ & $\sigma_\subin$ \\
        \hline
        \textbf{ESN}  & 400 & 1 & 1 & 99\% & 0.00176 & 12.2     \\ 
        \textbf{hESN} & 100 & 0 & 0 & 97\% & 0.25760 & 0.02825
    \end{tabular}
    \caption{Characteristics of the ESN and hESN. $\mathrm{sp}$ is the sparseness (i.e. fraction of 0 entries) of $\W$. The Tikhonov factor is $\gamma=10^{-9}$.}
\label{tbl:optimal_hyperparameters}
\end{table}

\subsubsection{Short-time prediction}
\label{sec:short-time_performance}

\begin{figure}[htbp!]
    \centering
    \includegraphics[width=\textwidth]{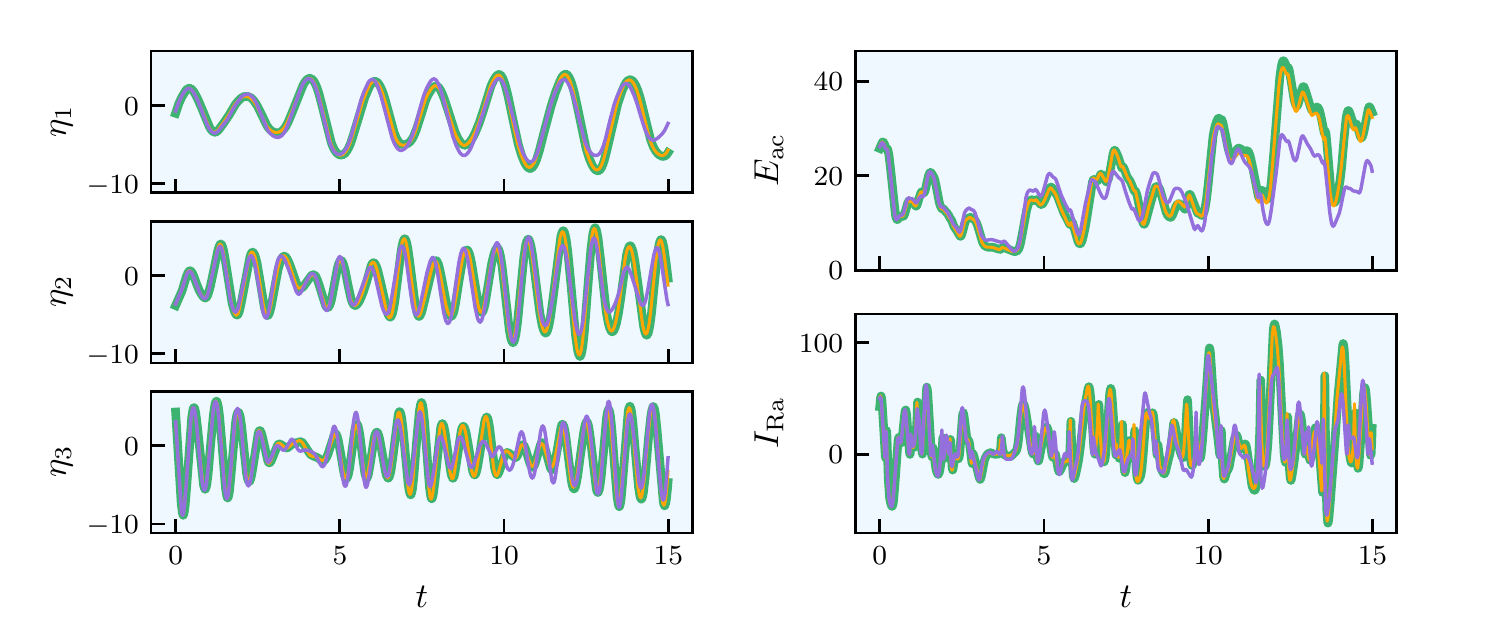}
    \caption{Short-time prediction. Time series of the first three (velocity) Galerkin modes, $\eta_1$, $\eta_2$ and $\eta_3$; acoustic energy, $\Eac$; Rayleigh index, $I_\mathrm{Ra}$.  
    }
    \label{fig:etas_Eac_IRa}
\end{figure}

\Cref{fig:etas_Eac_IRa} shows the time series of the first three (velocity) Galerkin modes, for the truth (data from ODE integration) and closed-loop predictions of the ESN and hESN. 
These modes are significantly more energetic than those of higher order because the flame is located at $x_f=0.2$, which markedly excites the first modes~\cite{Huhn2020a}.
All three modes oscillate in a non-periodic manner, with the peak frequency increasing with the mode number.
\Cref{fig:etas_Eac_IRa} also shows the time series of the two cost functionals, acoustic energy and Rayleigh index. 
The Rayleigh index oscillates substantially more than the acoustic energy because the time derivative of the acoustic energy is equal to the sum of the Rayleigh index and the dissipation from damping \cite{Huhn2020b}.
Time-wise, the hESN is able to time-accurately predict these modes for the whole time span shown, whereas the ESN deviates from the truth signal at $t \approx 10$.

\begin{figure}[htbp!]
    \centering
    \includegraphics[width=\textwidth]{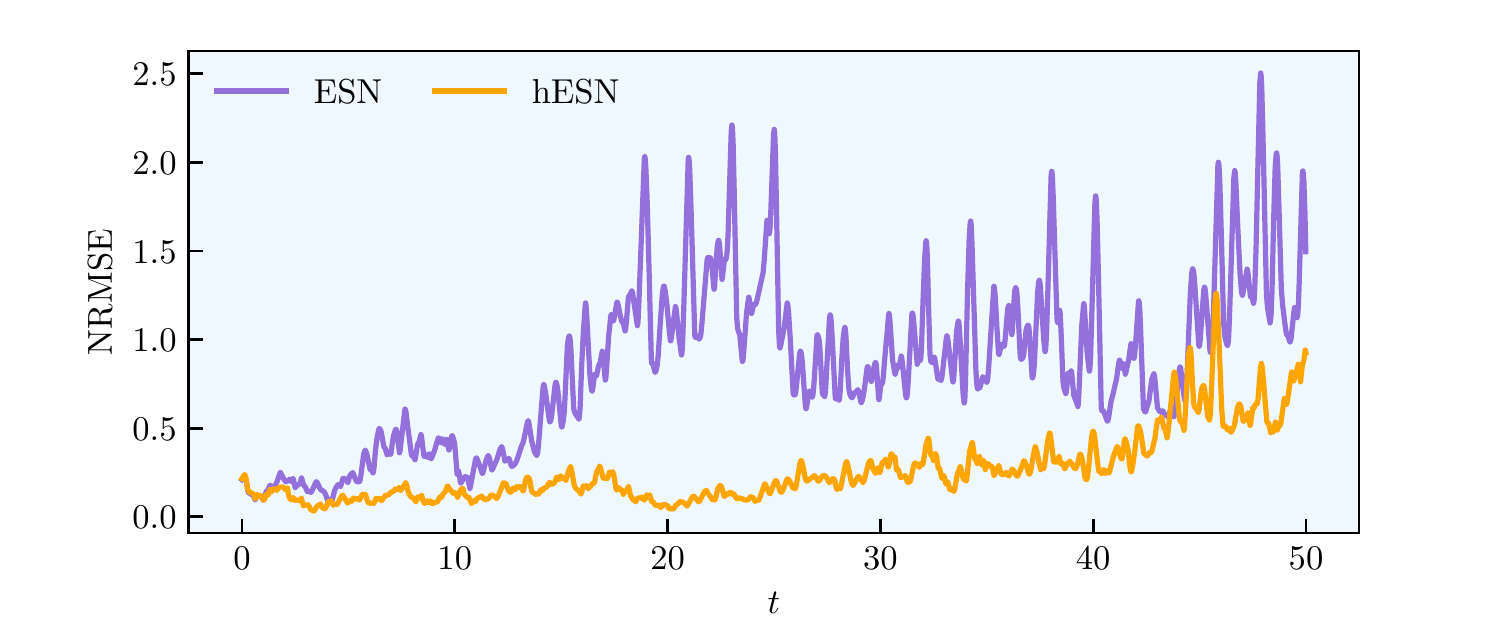}
    \caption{Short-time prediction. Normalized Root Mean Square Error of the acoustic modes}
    \label{fig:NRMSE}
\end{figure}

\begin{figure}[htbp!]
    \centering
    \includegraphics[width=\textwidth]{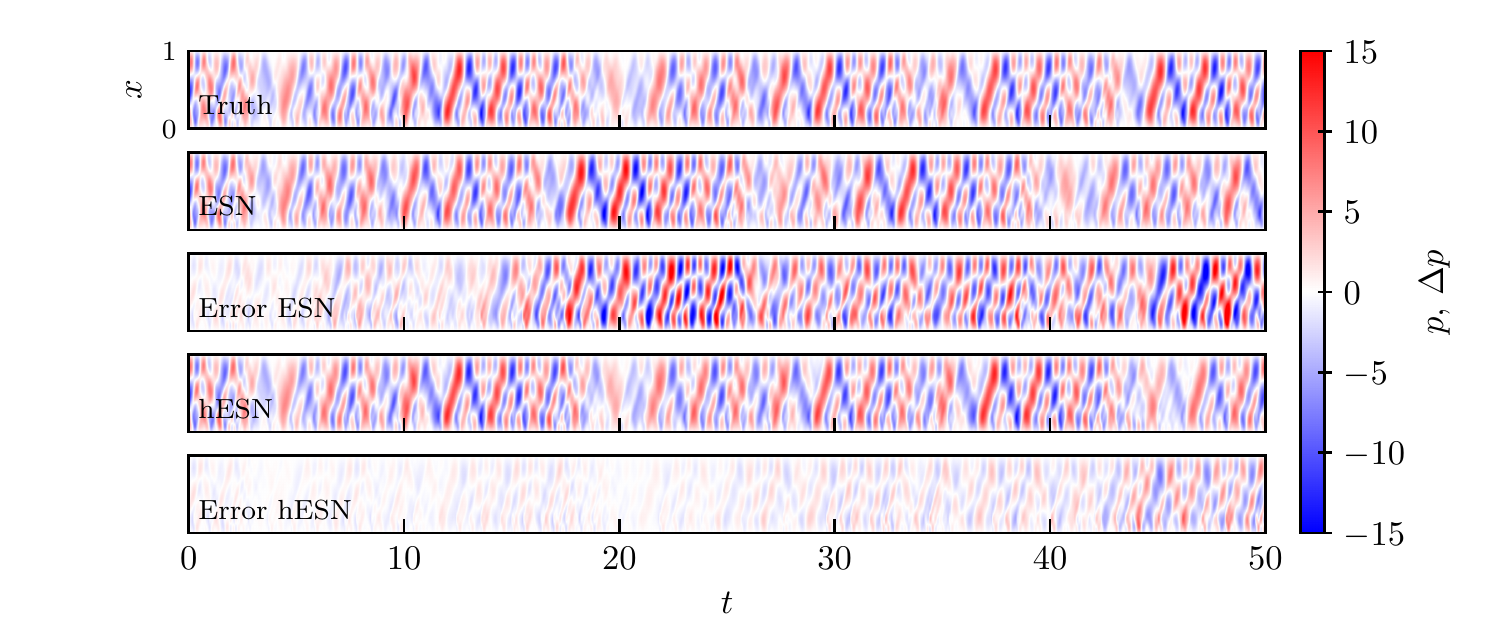}
    \caption{Short-time prediction. Acoustic pressure. The error is defined as $\Delta p(t,x) = \hat p(t,x) - p(t,x)$, where $\hat p$ is the prediction and $p$ is the true pressure field.}
    \label{fig:pressure_error}
\end{figure}

As a global metric, we compute the normalized root mean square error,
\begin{equation}
    \mathrm{NRMSE}(n) = \left(\frac{\left\vert\vert\hat{\bm y}(n) - \bm y(n)\vert\right\vert^2}{N^{-1}\sum_{j=1}^{N} \left\vert\vert\bm y(j)\vert\right\vert^2}\right)^{1/2},
\end{equation}
which is shown in \cref{fig:NRMSE}. 
The hESN performs better than the ESN. 
Given a threshold, the predictability horizon is defined as the time at which the first crossing of the threshold occurs~\citep{Pathak2018,Doan2019}. With a threshold of 0.5, the hESN achieves a predictability horizon of 42.1 time units (5.1 Lyapunov times, compared to 7.5 time units (0.9 Lyapunov times) of the ESN.
The NRMSE is, however, sensitive to normalization, and cannot discriminate between a time-inaccurate prediction that has similar dynamics and another that has completely different dynamics. 
%For example, let $\hat{\bm y}$ and $\bm y$ be two-dimensional. Then, if the first variable (i.e. first entries of $\hat{\bm y}$ and $\bm y$) is scaled by a factor $A$ and the second variable by $B$, the new NRMSE with the scaled variables will not be equal to the NRMSE without the scaling.
% 
%Furthermore, because of its global nature, it condenses the errors of different variables (potentially with different units) into one value.
%Due to this amalgamation of errors, the NRMSE does not have a clear physical interpretation.
%Furthermo, it also does not discriminate 
%
An alternative visualization of the short-time behavior is given in \cref{fig:pressure_error}, which shows the time evolution of the acoustic pressure in an $x-t$ diagram.
The truth panel shows that the flow is unsteady, featuring non-periodic acoustic waves propagating inside the domain. Although non-periodic, there appears to be a dominant frequency, showing up as roughly five waves in each 10 time units long window, which corresponds to an approximate period of 2 time units. This is related to the first acoustic eigenfunction. 
The third panel, corresponding to the ESN error, shows that the predictability horizon of the ESN is relatively short, which corroborates the findings of the NRMSE of \cref{fig:NRMSE}. 
However, the dynamics of the ESN are qualitative similar to those of the truth. 
This can be important, because, as shown in \cite{Huhn2020c}, an ESN can display inaccurate short-time prediction, but can have accurate long-time dynamics.
The findings from the pressure map agree with those of the NRMSE not only for the ESN, but also for the hESN. The pressure plot of the hESN indicates that it only starts to exhibit significant error at $t \approx 45$, which is similar to the  predictability horizon of 42.1 time units found with the NRMSE.
This result further shows that hESN is capable of time-accurate prediction.
We remark that such conclusions do not apply in general to the classes of conventional and hybrid echo state networks (i.e. an ESN need not perform worse than an hESN). 
Increasing the reservoir size of the ESN could yield satisfactory short-time prediction as well. 
However, the inclusion of model knowledge significantly improves the performance of an ESN for the same reservoir size.

\subsubsection{Long-time prediction}
\label{sec:long-time_performance}

% By inspecting the time evolution of both the NRMSE and the pressure field, we have investigated the time-accurate prediction capability of the ESN and hESN. 
% Notwithstanding, 
In this section, we focus the ergodic (i.e. long-time) prediction, which is key to this paper. 
As previously mentioned, inaccurate short-time (i.e. time-accurate) prediction does not necessarily imply inaccurate long-time prediction~\cite{Huhn2020c}. 
(Conversely, as shown in appendix~\ref{app:symmetric_lorenz}, accurate short-time prediction does not necessarily imply accurate long-time prediction either.) 
We analyse the predictive capability of the ESN and hESN in the long-time with metrics that are naturally defined in the statistically stationary regime. 

\begin{figure}[htbp!]
    \centering
    \includegraphics[width=0.5\textwidth]{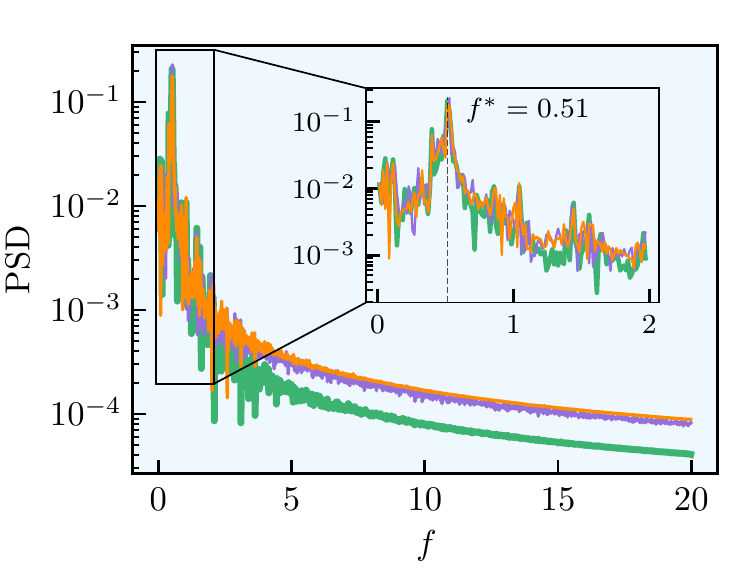}
    \caption{Long-time prediction. Frequency spectrum of the acoustic velocity mode, $\eta_1(t)$, for Truth, ESN and hESN.}
    \label{fig:frequency_spectra}
\end{figure}

First, we compute the frequency spectra (\cref{fig:frequency_spectra}). % 
The spectra are continuous, which is consistent with the underlying signal being chaotic.
The spectra match satisfactorily, with the largest error appearing at higher frequencies $f \gtrapprox 2$, which have a negligible importance because the power of the signal is concentrated at the lower frequencies. 
For lower frequencies (inset of \cref{fig:frequency_spectra}), there is a favourable agreement between the two types of networks and the truth.
In particular, both ESN and hESN match the dominant acoustic frequency and the peak of the true signal, which are close to the first natural acoustic eigenmode, $f=0.5$. This is consistent with the wave number in the $x$-$t$ plot of \cref{fig:pressure_error}.
Analysis of the spectra of the other state variables suggest similar conclusions (result not shown). 
We can conclude that echo state networks, both conventional and hybrid, reproduce the physical system satisfactorily in the time domain.

\begin{figure}[htbp!]
    \centering
    \includegraphics[width=\textwidth]{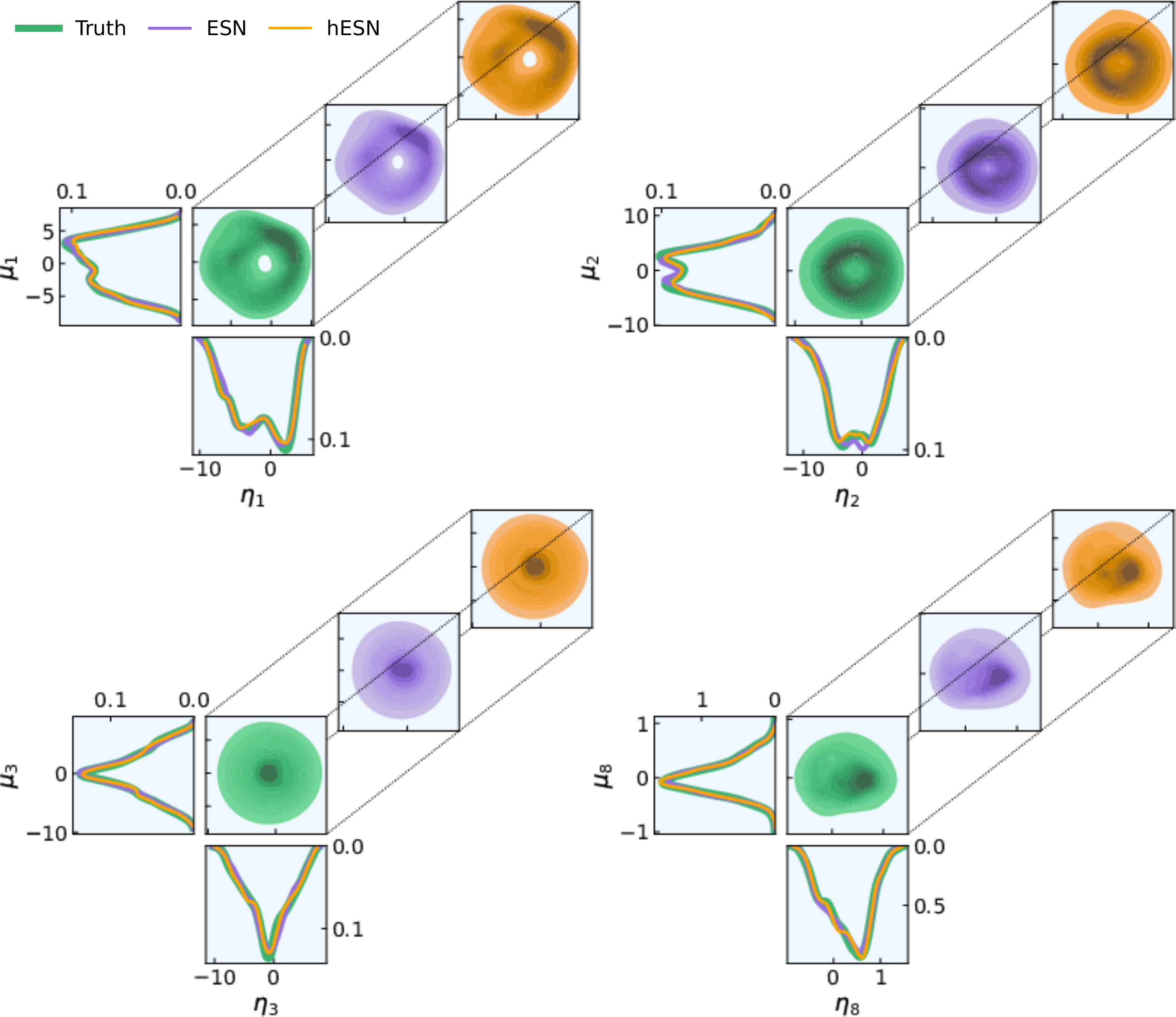}
    \caption{Long-time prediction. Probability Density Functions of the acoustic modes 1, 2, 3 and 8. The two-dimensional joint PDFs correspond to the velocity and pressure variables of the same acoustic mode pair, $(\eta_j, \mu_j)$. The one-dimensional PDFs are the marginalizations of the two-dimensional joint PDFs.}
    \label{fig:PDFs}
\end{figure}

Second, we compute the probability density functions (PDFs) of the chaotic attractor because the long-time behavior of a system and its statistics depend on the invariant measure of its attractor.
%Thus, the natural comparison of the long-time behaviors between the truth and the networks is the comparison of between their attractor probability density function (PDFs).
%Unfortunately, due to the high dimensionality of the system, that is unfeasible \fh{(Unfeasible is not the best word. I know there's a better one, but I can't think of it atm.)}. 
%
We compute two-dimensional joint PDFs of the Galerkin modes, i.e. $(\eta_1, \mu_1)$, $(\eta_2, \mu_2)$, etc.; and the one-dimensional PDFs of the individual state variables. These are performed for modes 1, 2, 3 and 8 (\cref{fig:PDFs}), the first three being the most energetic and the last being representative of the higher modes. The PDFs are obtained via Kernel Density Estimation~\cite{Scott1992,Scipy2020}. 
Both networks perform well, with their PDFs matching those of the truth relatively well.
However, although the ESN shows a good agreement with the truth, it is less accurate than the hESN.  
The difference in performance is more evident close to the modes of $\eta_1$ and $\eta_2$, where the ESN over- and underpredicts the values of the peaks.
This indicates that the invariant measure of the attractor of the dynamical system is well captured by both the ESN and hESN, with the latter being  more accurate.
Therefore, both ESN and hESN  predict the long-time statistics of the physical system, with training from relatively small data. 

\subsection{Design optimization}
\label{sec:chain_optimization}

For the physical optimizer, we define the cost functional, $\avg{\Eac}$; the optimization space, $\beta \in [7.5, 10.0], \, \tau \in [0.1, 0.3]$; the acquisition function, LCB with $\kappa=1.960$; the covariance function, RBF; the number of initial seed points, 4, which should be sufficient to properly initialize the GP; and the maximum number of evaluations (12, including the seed points), which we find to be a good compromise between efficacy and efficiency\footnote{Efficacy here relates to whether the goal is achieved or not, whereas efficiency relates to how costly achieving the goal was.}.
Similarly, for the hyperparameter optimiser, we define the cost functional, validation MSE; optimisation space, $\log_{10}(\sigma_\subin) \in [-2, 2]$, $\log_{10}(\rho) \in [-3, 0]$, $\log_{10}(\gamma) \in [-11, -4]$; the acquisition function, GP-Hedge; the covariance function, Matérn 3/2; the number of initial seed points, 5; and the maximum number of evaluations, 25, which include the seed points.
Here, we include the Tikhonov factor as a hyperparameter.
The reason is that the physical optimization evaluates attractors that can vary widely.
In this case, adjusting the Tikhonov factor becomes important because it controls the relative importance of the MSE and the norm of $\W_\subout$ in the training problem, two terms that can vary substantially depending on the attractor.
This is in contrast with \Sref{sec:performance}, where only one attractor was learned and predicted.
To save on computational cost, the hyperparameter optimization stops when the error is below the threshold of $3 \times 10^{-2}$ (\Sref{sec:hyperparameter_selection}).
The chain then runs on its own.
%
%%%%%%%%%%%%%%%%%%%%%%%%%

First, the physical optimizer randomly generates seed points.
For each of these points, $\bm s_\mathrm{phys} = (\beta, \tau)$, data, $\{\bm q(0), \bm q(1), \dots\}$\footnote{Whereas $\bm q$ is defined in \cref{eq:ode} as continuous in time, here, $\bm q$ is the numerical solution, which is only defined at discrete times $0, 1, \dots$. Thus, slightly abusing notation, we write $\bm q(t=n\, \Delta t)$ as $\bm q(n)$, where $n$ is a discrete time.}, is generated by integrating \cref{eq:rijke_discrete1,eq:rijke_discrete2,eq:kings_law}.
Then, the hyperparameter optimizer selects the optimal hyperparameters, $\bm s_\mathrm{hyper}^+ = (\sigma_\mathrm{in}^+, \rho^+, \gamma^+)$, using Bayesian optimization. % , inside which the network is trained in open-loop (\cref{eq:hesn_r,eq:hesn_y,eq:Wout}.
With the optimal hyperparameters (and the corresponding optimal $\Wout$), the network is run in closed-loop, generating a long time series, $\{\hat{\bm y}(0), \hat{\bm y}(1), \dots\}$, from which the time-averaged acoustic energy, $\avg{\Eac}$, is computed and returned to the physical optimizer.
After the seed points have been evaluated, the physical optimizer selects the next point by finding the optimum of the acquisition function, $\bm s_\mathrm{phys} = (\beta, \tau)$, which is then evaluated in the same manner as the seed points. 
The optimisation stops when 12 points (including seed points) have been evaluated.

\begin{figure}[tb]
    \centering
    \includegraphics[width=\textwidth]{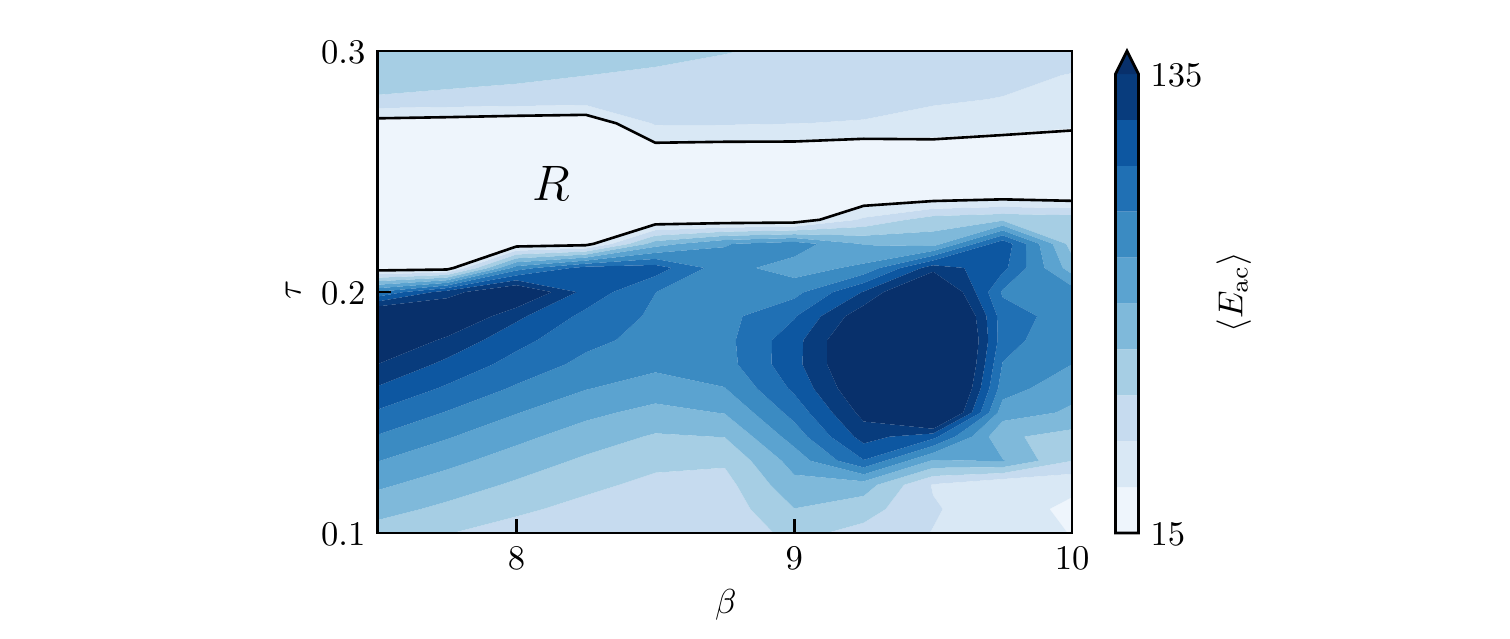}
    \caption{Time-averaged acoustic energy, $\langle \Eac \rangle$, versus the flame parameters, $\beta$ and $\tau$,  obtained with a brute-force grid search. Benchmark solution.}
    \label{fig:2d_obj_grid}
\end{figure}

For comparison, the true cost functional is physically shown in \cref{fig:2d_obj_grid}.
This chart is generated by integrating the ODEs on a grid of 11 values of $\beta$ and 21 of $\tau$. 
There is a large region of high acoustic energy, which can be divided into two sub-regions, each centered around a local maximum, one of which is on the boundary of the domain ($\beta=7.5$ and $\tau \approx 0.18$).
Above this, there exists a nearly horizontal strip spanning the whole range of $\beta$, marked $R$ in \cref{fig:2d_obj_grid}, which is where the optimum from the optimisation is likely to be found. Physically, as noted in~\citet{Huhn2020a}, the time-averaged acoustic energy may be discontinuous at certain flame parameters because the attractor is structurally unstable. 
The global minimum found with the grid of \cref{fig:2d_obj_grid} is $\avg{\Eac}(\beta \approx 8.25, \tau \approx 0.27) = 15.04$.

%%%%%%%%%%%%%%%%%%%%%%%%%%%%%%%%%%%%

\begin{figure}[htb]
    \centering
    \includegraphics[width=\textwidth]{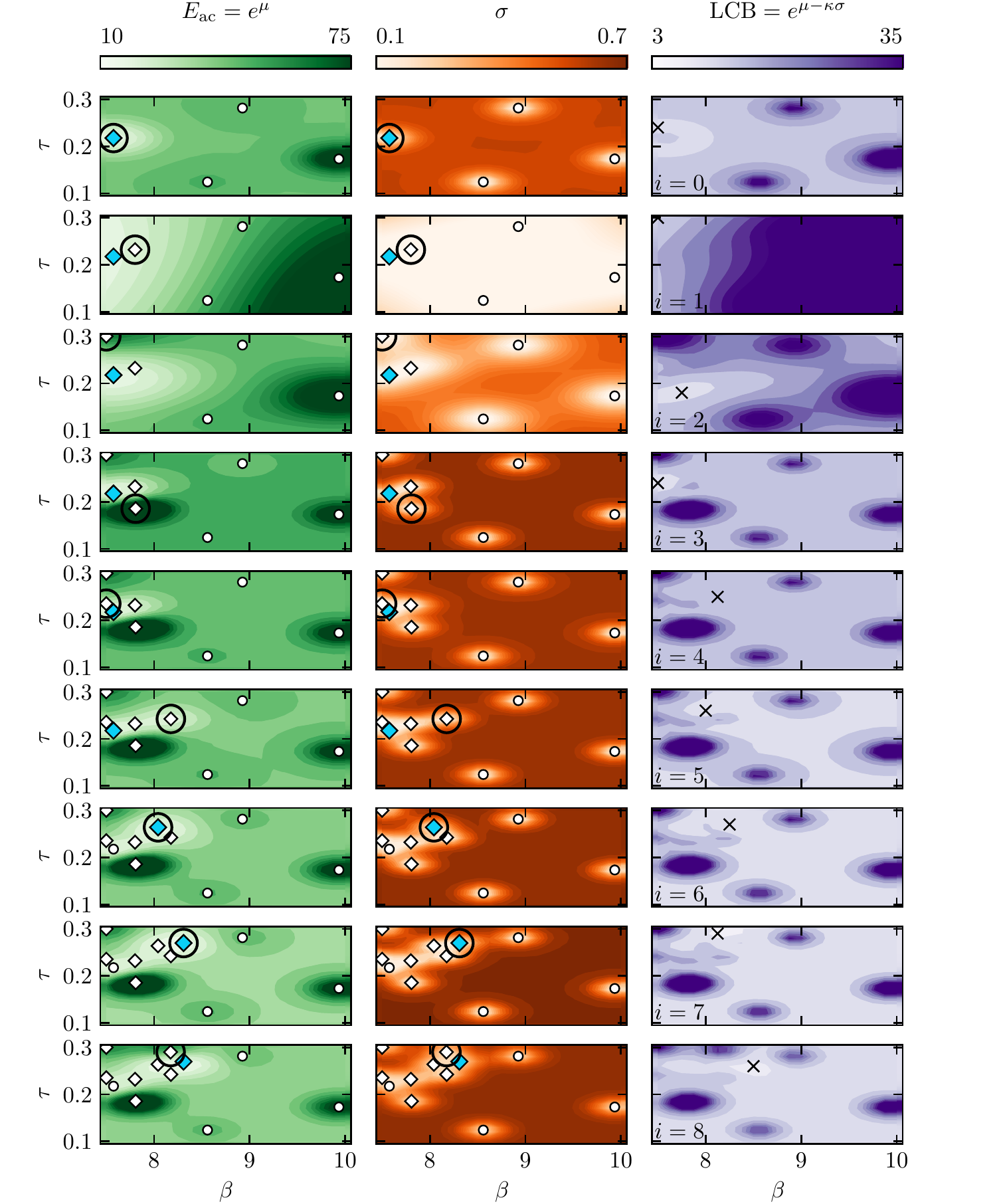}
    \caption{Optimisation history. The first and second columns show the mean and standard deviation of the GPR. Seed points and previously evaluated points are marked with white circles and diamonds, respectively. The current optimum is blue and the last evaluated point is encircled. The third column shows the acquisition function, LCB, the minimum of which is the next point to be evaluated (cross).}
    \label{fig:2d_optim_history}
\end{figure}

\begin{figure}[htb]
    \centering
    \includegraphics[width=\textwidth]{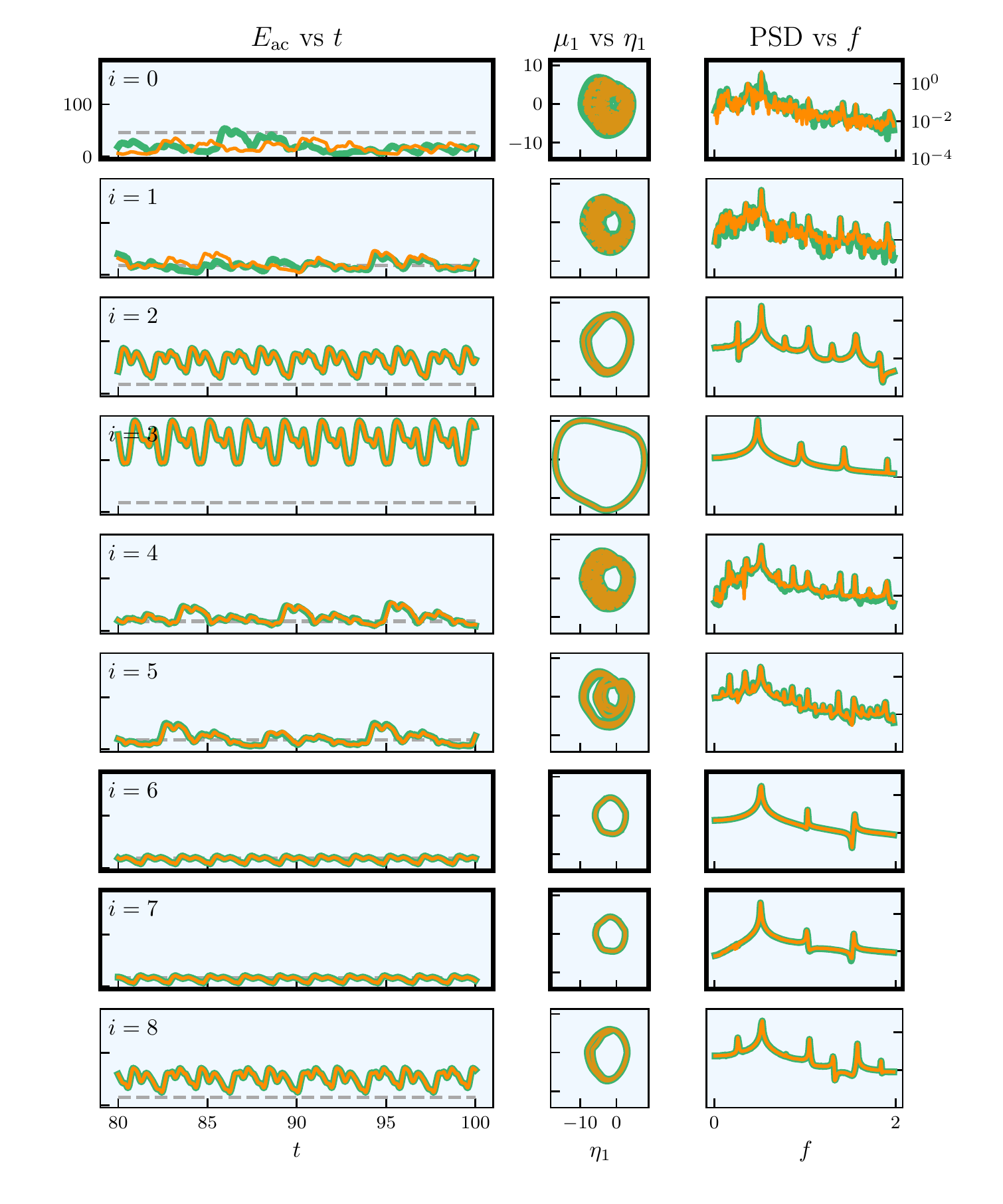}
    \caption{Evaluated points during optimisation. The columns correspond to the last fifth of the time series of the acoustic energy; the phase plot of the first acoustic mode ($\mu_1$ vs $\eta_1$); and the frequency spectrum of the first acoustic velocity mode, $\eta_1$. Rows with thick spines correspond to a new optimum. The time-averaged acoustic energy of the current optimum is shown for reference as a horizontal dashed grey line.}
    \label{fig:optim_ts_pp_psd}
\end{figure}

\Cref{fig:2d_optim_history} shows the results of the physical optimisation. 
The three columns correspond to the mean of the GPR, standard deviation of the GPR, and the acquisition function LCB.
The $i$-th row (starting at 0) corresponds to the state of the optimisation after $n_\mathrm{seed}+i$ evaluations, where $n_\mathrm{seed}$ is the number of initial seed points (4 here) used to seed the optimization. 
It shows the previously evaluated points, with the most recent being encircled. 
The minimum of the acquisition function, and therefore the next point to be evaluated, is marked with a cross in the third column.
Each row in \cref{fig:2d_optim_history} corresponds to a row of \cref{fig:optim_ts_pp_psd}, which contains the time series of the acoustic energy, the phase plot $\mu_1$ vs $\eta_1$ and the frequency spectrum of $\eta_1$ of the newly evaluated point. For the purpose of comparison, we include the true signals.

Initially (row $i=0$), with 4 seed points, the GPR indicates a two-dimensional dependence on both $\beta$ and $\tau$, with clear regions of similar acoustic energy around each of the seed points, especially the two at the extremes of $\beta$.
This is because these two points correspond to low and high values of acoustic energy.
The combination of a larger distance in $\beta$ than in $\tau$ between these two points, and the fact that the other two points, which have low and high values of $\tau$, have similar $\avg{\Eac}$, leads the fit of the GPR to place a larger weight on $\beta$ than $\tau$.
In terms of dynamics, the optimum of the seed points is a chaotic attractor, as shown in the first row of \cref{fig:optim_ts_pp_psd}.

The agreement between truth and hESN is favorable. Furthermore, the time series remains below the dashed line, which corresponds to the previous optimum, showing that this design is the optimum of the seed points.
As expected, the uncertainty is lower in regions centred around evaluated points.
The dependence on $\tau$ is substantially reduced after the evaluation of the first selected point ($i=1$), a design that is close to the current optimum, not only in distance in the design space, but also in time-averaged acoustic energy (\cref{fig:optim_ts_pp_psd}).
The perceived small dependence on $\tau$ and relatively strong dependence on $\beta$, in conjunction with low estimated uncertainty, makes the acquisition function discard approximately three quarters of the optimization domain, corresponding approximately to the upper three quarters of the range of $\beta$ (\cref{fig:2d_optim_history}, $i=1$, third column).
With uncertainty low almost everywhere, its highest values are found close to the corners of the design space, where the distance from the sampled points is maximal.
Moreover, given that the GPR indicates positive dependence of the cost functional on $\beta$ and slightly negative dependence on $\tau$, the minimum of the acquisition function is naturally found at the upper left corner, where $\beta$ is minimal and $\tau$ maximal.
This point ($i=2$), in contrast with the estimate prior to its evaluation, is in a region of moderately-high acoustic energy (\cref{fig:2d_obj_grid}). This can be verified in \cref{fig:optim_ts_pp_psd}, which shows that this combination of $\beta$ and $\tau$ corresponds to a limit cycle whose instantaneous acoustic energy never drops below the time-averaged acoustic energy of the current optimal design.
The new data updates the GPR, which now exhibits moderate uncertainty throughout the domain, except relatively close to previously evaluated points ($i=2$ row, second column).
While the dependence on $\beta$ remains, the dependence on $\tau$ increases and is no longer monotonic.
With moderately-high uncertainty away from the points, and with low mean in a small region only around the current optimum, the acquisition function selects a point where the two regions (high uncertainty, low mean) ``meet''.
The new design point, however, has very high acoustic energy (row $i=3$ of \cref{fig:optim_ts_pp_psd}), as it belongs to the area surrounding the leftmost of the two peaks of high acoustic energy (\cref{fig:2d_obj_grid}).
This newly acquired information strongly contrasts with the prior estimate of the GPR. Naturally, this is because of the sharp variation of acoustic energy at the lower edge of the region $R$. A small variation of parameters from the current optimum resulted in high variation of the cost functional. As such, uncertainty is now high everywhere, except close to previously sampled points.
At $i=4$, the acquisition function chooses a point close to the current optimum, which is a clear evidence of exploitation. This design results in chaotic dynamics, similarly to the current optimum, but it does not produce lower time-averaged acoustic energy. Thus, it is not a new optimum.
With three points (and the left boundary) enclosing the current optimum, there is little advantage in continuing exploiting. Switching to exploration, the new point ($i=5$) is relatively far from the current optimum. Once again, it is at the intersection of low mean and high uncertainty, since that combination minimises the acquisition function. While the new design does not improve the optimum, it does provide crucial new information. Because its acoustic energy is low, the region of low mean expands with the updated GPR.
This new expansion provides space to exploit. Thus, a new design ($i=6$) relatively close to the previous is selected. This new design offers lower time-averaged acoustic energy than the current optimum, i.e. the optimization found a new optimum. The new optimal design represents an improvement of 8.4\% with respect to the previous optimum. In the GPR, not only has a new optimum been found, but the region of low mean is now larger.
Thus, at $i=7$, the most recently selected design further expands the spread of points into higher $\beta$ and higher $\tau$. Similarly to the last design, a new optimum is found, this one offering a further 11.4\% reduction of acoustic energy.
Finally, the last design ($i=8$), despite being close to the previous two optima, does not further improve the cost functional.  It is likely that this design is above the upper boundary of the region $R$ in \cref{fig:2d_obj_grid}.
Had the optimisation continue to run, it is possible the optimum could be slightly improved.
However, the maximum number of evaluations was reached.
Furthermore, the optimum of the optimization, found with 12 design evaluations (4 seed points plus 8 selected points), $\avg{\Eac}(\beta\approx 8.31, \tau\approx 0.27) = 14.68$, is slightly better than that found with a brute-force grid search (\cref{fig:2d_obj_grid}), $\avg{\Eac}(\beta\approx 8.25, \tau\approx 0.27) = 15.04$, which needed 231 evaluations. 
A larger number of evaluations would likely have been a relatively poor trade-off between design improvement (i.e. decrease in cost functional) and computational cost.

\begin{figure}[tb]
    \centering
    \includegraphics[width=0.5\textwidth]{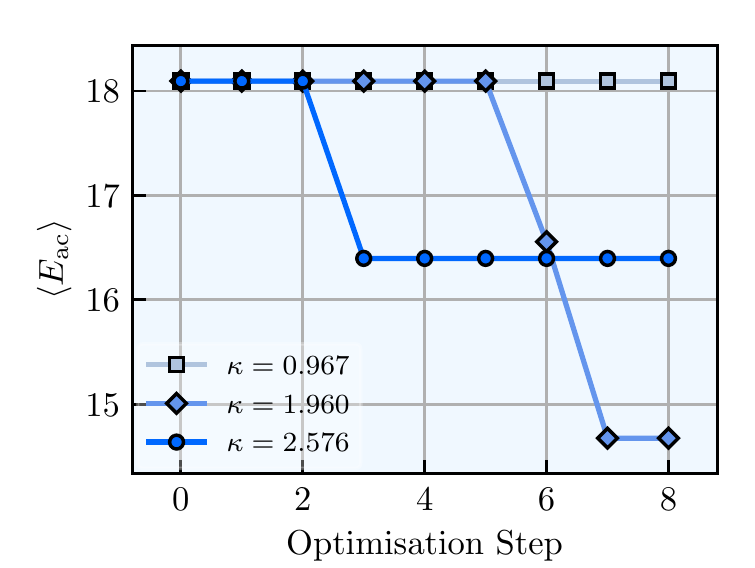}
    \caption{Time-averaged acoustic energy, $\langle \Eac \rangle$, versus number of points evaluated, for three values $\kappa$: 0.967, 1.960 and 2.576; corresponding to 67, 95 and 99\% confidence intervals.}
    \label{fig:2d_optim_convergence}
\end{figure}

\Cref{fig:2d_optim_convergence} shows the convergence of the optimisation procedure, i.e. the current optimum versus the number of points evaluated for three values of $\kappa$.

\begin{figure}[tb]
    \centering
    \includegraphics[width=\textwidth]{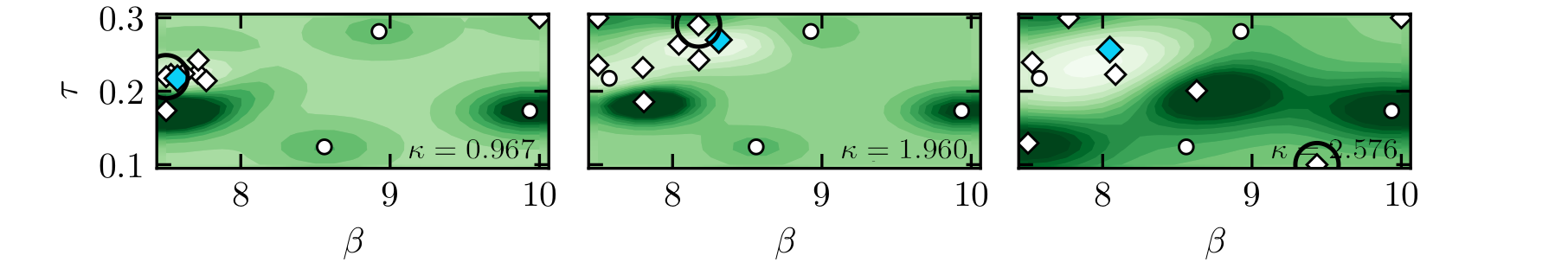}
    \caption{Final state of optimisation with three values of $\kappa$: 0.967, 1.960 and 2.576; corresponding to 67, 95 and 99\% confidence intervals. This shows the effect of $\kappa$ in the balance between exploration vs exploitation.}
    \label{fig:kappas_parametric}
\end{figure}

The largest value of $\kappa$, $2.576$, favours exploration the most. This results in quickly, in the second optimisation step, finding a design that improves on the best seed point. However, because of its tendency to explore, it does not try to exploit and locally improve its current optimum as much.
Hence why there is a large spread of points for $\kappa=2.576$ in \cref{fig:kappas_parametric}, which shows the last state of the optimisation for the same three values of $\kappa$ of \cref{fig:2d_optim_convergence}.
% Perhaps, had the optimisation continued to run, more exploitation would have occurred, but necessarily at higher computational cost.
%
In contrast, $\kappa=0.967$, the lowest of the three, will seek to mostly exploit. The various designs concentrated in a small region are evidence of this. Unsurprisingly, this does not produce a better design that the optimal seed point.
Finally, $\kappa=1.960$, used in the optimisation of \cref{fig:2d_optim_history,fig:optim_ts_pp_psd}, navigates between these two lines, exploration and exploitation, unsuccessfully trying to exploit initially, finding a new optimum with exploration and subsequently improving the recent optimum by exploiting its surrounding region.
%These effects can be seen in \cref{fig:kappas_parametric}, which shows the mean of the last GPR of each optimisation.
In conclusion, the larger $\kappa$ is, the larger the spread of points, which shows the influence that this parameter has on the balance between exploration and exploitation.
In this optimization problem, the initially chosen value of $\kappa=1.960$ seems to offer the best performance among the three.

\section{Conclusions}

Gradient-based design optimization of chaotic acoustics is notably challenging for a threefold reason.  
First, first-order perturbations grow exponentially in time, which makes the computation of the gradients with respect to the design parameters ill-posed. 
Second, the statistics of the solution may have a slow convergence, which makes the time integration of the equations computationally expensive. 
Third, chaotic acoustic systems may have discontinuous variations of the time-averaged energy~\citep{Huhn2020a}, which means that the gradient may not exist for all design parameters.
In this paper, we develop an optimization method to find the design parameters that minimize time-averaged acoustic cost functionals.  
The method is gradient-free, with Bayesian sampling; model-informed, with a reduced-order acoustic model; and data-driven, with reservoir computing. 

First, we analyse the predictive capabilities of reservoir computing based on echo state networks. 
Both fully data-driven and model-informed architectures are considered.
In the short-time prediction, model-informed networks can time-accurately predict the chaotic pressure oscillations beyond the predictability time (the Lyapunov time).
For the same reservoir size, informing the training with a cheap model extends the prediction from  $\sim1$ Lyapunov time to  $\sim5$ Lyapunov times.   
In the long-time prediction, we show that echo state networks accurately reproduce the statistics of chaotic acoustic attractors. 
The hyperparameters are automatically tuned by using Bayesian optimization, which provides a consistently good performance across different architectures, reservoir sizes and data. 
With accurate predictions at a  lower computational cost, the long-time series are generated to obtain the time-averaged acoustic energy that is being optimized. 
%Furthermore, we showed that by supplementing the purely data-driven echo state network with physical knowledge, the quality of the predictions can be substantially improved, leading to reduced network sizes.
Second, we couple echo state networks with a Bayesian technique based on Gaussian Processes to explore the design parameter space. 
%by coupling a gradient-free optimization algorithm with the procedure of generating a small amount of training data, automatically tuning the hyperparameter via Bayesian optimization and obtaining statistics from a long time series generated by a network. 
The computational method is minimally intrusive because it requires only the initialization of the physical and hyperparameter optimizers; e.g., a factor to balance the exploration versus exploitation of the sampling; and the reservoir size.
Third, we apply the computational method to the minimization of the  time-averaged acoustic energy of chaotic oscillations. 
We focus on the acoustics that is excited by a heat source, which is relevant to thermoacoustic oscillations in propulsion and power generation. 
The design parameters that are changed during the optimization are the flame intensity and time delay. Nonetheless, the method can tackle other physical parameters.
Starting from five random designs with energetic chaotic oscillations, we find an optimal set of parameters in 8 iterations. This optimum is practically equal to the optimum found by brute-force grid search, which needs 231 evaluations.
%\lm{[Can you add a couple of physical remarks here? We need this for PRF.] We find that ... }
The thermoacoustic system shows a variety of solutions and bifurcations during the optimization update (e.g., limit cycles, strange attractors), which are accurately learnt by the echo state networks. This is because the echo state network learns the physical temporal correlations of the acoustic modes through the sparse recurrent dynamics of the reservoir. 
%\fh{In the process of optimizing the system, both periodic and chaotic oscillations emerge. Yet, the echo state network used was capable of predicting either type of solution, showing the robustness}

This work opens up new possibilities for the optimization of chaotic systems, in which the cost of generating data, for example from high-fidelity simulations and experiments, is high. 
\section*{Declaration of Interests}
The authors report no conflict of interest. 

\section*{Acknowledgements}
F. Huhn is supported by Fundação para a Ciência e Tecnologia under the Research Studentship No. SFRH/BD/134617/2017. L. Magri gratefully acknowledges financial support from the ERC Starting Grant PhyCo  949388, and TUM Institute for Advanced Study (German Excellence Initiative and the EU 7\textsuperscript{th} Framework Programme n. 291763). The authors thank A. Racca for insightful discussions.

\clearpage 

\appendix
\section{Divergence of hybrid echo state network}
\label{app:divergence}

\begin{figure}[htbp!]
    \centering
    \includegraphics[width=0.5\textwidth]{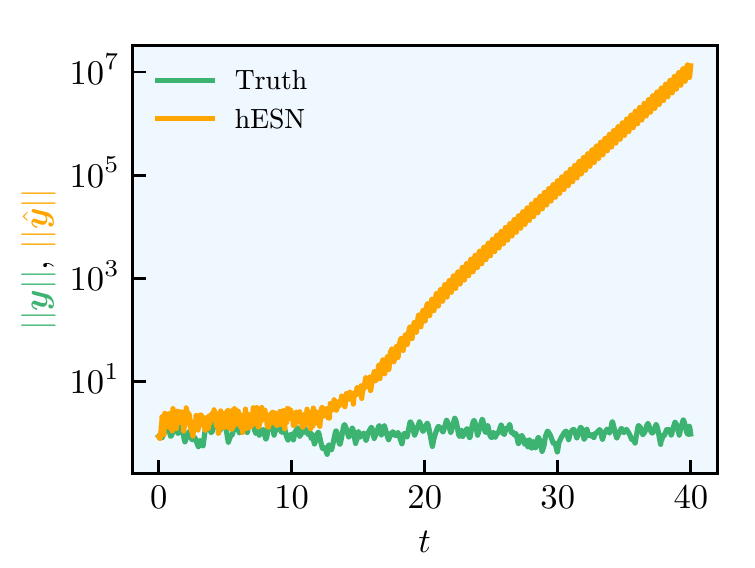}
    \caption{Divergence of the prediction of a hybrid echo state network. This is obtained with the configuration detailed in \Sref{sec:hyperparameter_selection} and $\sigma_\subin = 4.23$ and $\rho = 0.9$. Other combinations of hyperparameters and physical parameters can result in similar behavior.
    }
    \label{fig:hybrid_divergence}
\end{figure}

Unlike conventional echo state networks, whose output is bounded (though the bound can be very far from the attractor), due to the feedback from the output of $\bm \K$ to its own input via $\Wout$, hybrid echo state networks can diverge to infinity. This can be seen in \cref{fig:hybrid_divergence}. 
However, it should be noted that this behavior is not necessarily a function of the physical parameters or hyperparameters only. A certain fixed combination of hyperparameters can result in both divergence and non-divergence depending on the physical parameters. Similarly, for fixed physical parameters, changing the hyperparameters can result in divergence or not.
This is not an issue with the training method (ridge regression), but the complex combination of training and validation time series; realizations of $\Win$ and $\W$; values of $\sigma_\subin$ and $\rho$; Tikhonov factor $\gamma$; model $\K$ and its numerical scheme; etc.
Furthermore, $\K$ can be stable (i.e. its solution is stable), but the hybrid echo state network using it can be unstable. In fact, that is the case of \cref{fig:hybrid_divergence}, where, if $\K$ evolved on its own, a limit cycle arises.
It is the (linear) transformation due to the output weights $\Wout$ that changes the output of $\K$, which is then fed back to $\K$ itself, that can make the whole system unstable.

A very succinct example, where we omit the reservoir nodes for simplicity, is
\begin{equation}
    \mathcal{K}(y) = (1-\lambda) y,
\end{equation}
where $0<\lambda<1$ is a physical parameter. If left to evolve on its own, i.e.
\begin{equation}
    y(n+1) = \mathcal{K}(y(n)) = (1-\lambda) y(n),
\end{equation}
$y$ will converge to 0. However, in the hybrid echo state network framework, we have instead
\begin{equation}
    y(n+1) = \mathcal{K}(W_\subout y(n)) = W_\subout (1-\lambda) y(n).
\end{equation}
For $y$ to converge to 0, we must have
\begin{equation}
    W_\subout < \frac{1}{1 - \lambda}.
\end{equation}
If $W_\subout$ does not verify this condition, the system will diverge, despite $\mathcal{K}$ being stable.

\section{Accurate short-time, inaccurate long-time prediction}
\label{app:symmetric_lorenz}

\begin{figure}[h]
     \centering
     \begin{subfigure}{0.48\textwidth}
        \includegraphics[width=\textwidth]{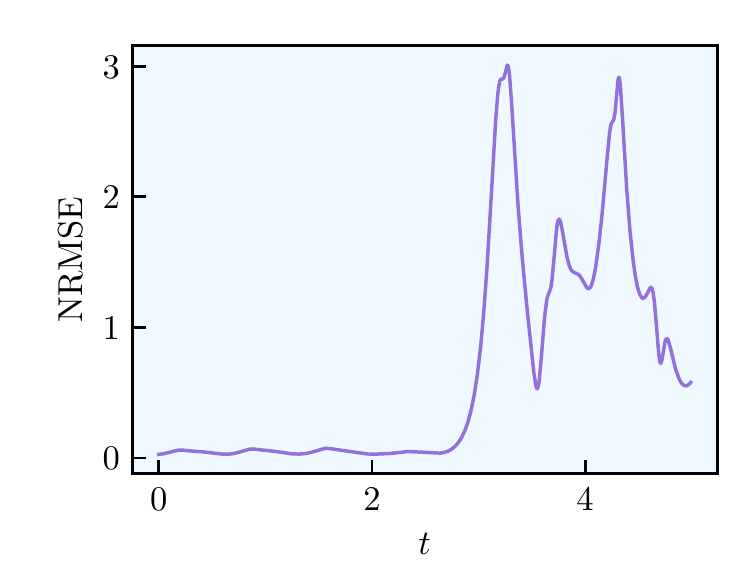}
     \end{subfigure}
     \begin{subfigure}{0.48\textwidth}
        \includegraphics[width=\textwidth]{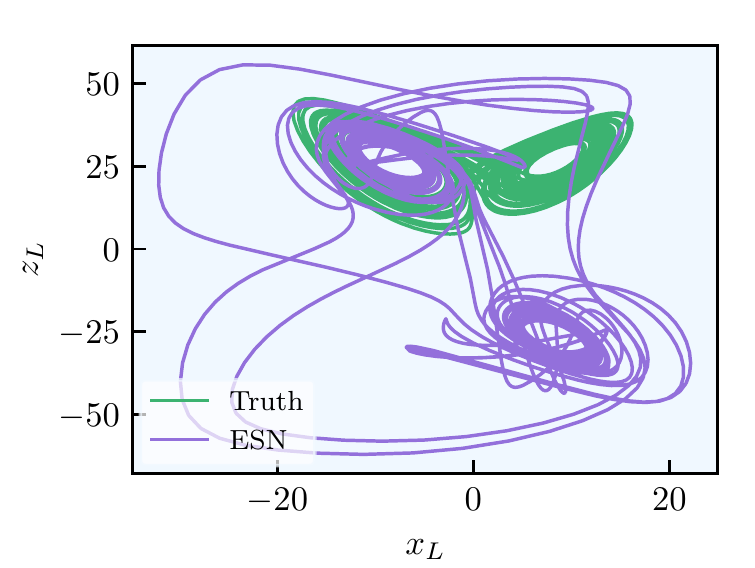}
     \end{subfigure}
     \caption{Lorenz system. Time series of NRMSE and phase plot $z_L$ vs $x_L$.}
     \label{fig:lorenz}
\end{figure}

Here, we use a short example based on the Lorenz system~\citep{Lorenz1963} to show that accurate short-time prediction does not necessarily mean accurate long-time prediction.
The Lorenz system is a three-dimensional system,
\begin{align}
    \frac{dx_L}{dt} &= \sigma_L (y_L - x_L), \\
    \frac{dy_L}{dt} &= x_L (\rho_L - z_L) - y_L, \\
    \frac{dz_L}{dt} &= x_L y_L - \beta_L z_L,
\end{align}
where $\sigma_L$, $\rho_L$ and $\beta_L$ are parameters, often equal to 10, 28, 8/3, which is a combination that produces chaotic motion.
This system is numerically integrated with time step 0.01 to generate training, validation and test data.
For this example, we use an echo state network with 100 nodes with no biases. The network is trained and validated on datasets of length 500, using Bayesian optimization.
\Cref{fig:lorenz} shows the NRMSE and (long-time) phase plot for the Lorenz system.
The NRMSE remains below the threshold of 0.2 until $t \approx 2.87$, which corresponds to approximately 2.6 Lyapunov times (leading Lyapunov exponent of approximately 0.9). Thus, the ESN predicts the system relatively well in the short time.
However, the phase plot shows a completely different behavior between prediction and data in the long-time.
In this case in particular, the network has no biases (i.e. $b_\subin = b_\subout = 0$), in which case the reservoir evolves according to
\begin{equation}
    \bm r(n) = \tanh(\widetilde{\bm W} \bm r(n-1)),
\end{equation}
where $\widetilde{\bm W} = \W + \Win\Wout$~\citep{Huhn2020c}. This means that taking some reservoir state $\bm r(n-1)$, and flipping its sign, i.e. $\bm r'(n-1) = -\bm r(n-1)$, one gets $\bm r'(n) = -\bm r(n)$. Thus, either the ESN admits two attractors symmetric of each other, or admits one symmetric attractor, which is the case here.
In conclusion, accurate short time prediction does not necessarily imply accurate long-time prediction.

\section{\fh{Computational cost of the method}}
\label{app:cost_analysis}

\fh{
The optimization framework in this paper (i.e. the ``chain'') was demonstrated on a relatively low-dimensional system. In this particular case, the echo state network (and everything it involves) could have been foregone and Bayesian optimization applied directly to the result of the (longer run) ODEs. However, the application in this work is a proof of concept and not an example of computational gains. These are meant to be achieved in larger-scale systems, such as high-fidelity simulations.
The cost of the method is
\begin{equation}
    N_\mathrm{opt} (N_\mathrm{train} C_\mathrm{ODE} + C_\mathrm{train} + N_\mathrm{test} C_\mathrm{ESN}),
\end{equation}
where $N_\mathrm{opt}$ is the number of optimization steps in the physical domain, $N_\mathrm{train}$ is the number of training timesteps, $C_\mathrm{ODE}$ is the cost per timestep of solving the ODEs, $C_\mathrm{train}$ is the cost of training the network (including the hyperparameters), $N_\mathrm{test}$ is the number of test timesteps and $C_\mathrm{ESN}$ is the cost of per timestep of the closed-loop ESN.
On the other hand, applying Bayesian optimization directly to the ODEs instead would have a cost of
\begin{equation}
    N_\mathrm{opt} N_\mathrm{test} C_\mathrm{ODE}.
\end{equation}
For there to be a cost benefit
\begin{equation}
     N_\mathrm{opt} (N_\mathrm{train} C_\mathrm{ODE} + C_\mathrm{train} + N_\mathrm{test} C_\mathrm{ESN}) < N_\mathrm{opt} N_\mathrm{test} C_\mathrm{ODE}
     \label{eq:cost_analysis}
\end{equation}
must be verified.
Thus, each term must be analyzed. The most expensive operation is training the network, $C_\mathrm{train}$, which scales with $O(N_x^3)$ due to the matrix inversion. If the number of nodes, $N_x$, is proportional to the dimension of the dynamical system, $N_d$, then this cost becomes $O(N_d^3)$. This operation only happens once (per hyperparameter combination), though.
Once it is performed, the largest cost is the closed-loop ESN simulation at $O(N_x N_d) \sim O(N_d^2)$ per timestep.
On the other hand, the cost of a timestep in a numerical simulation (ODE) can scale with $O(N_d^2)$ or $O(N_d^3)$, depending on the numerical scheme. This would put it at a similar scaling to the networks.
Therefore, it would seem hard for \cref{eq:cost_analysis} to be verified.
However, the goal is not to apply the technique directly to a high-fidelity simulation, but to a lower resolution of its results. While a certain number of grid points may be needed for an accurate simulation, after obtaining the results, only a subset of these points is required for the accurate computation of the cost functional. In other words, not all points are needed. Thus, the full state vector need not be computed/predicted. For example, if there is a downsample of 10 to 1 in every direction of a 3D high-fidelity simulation, there is a $1000$-fold reduction in $N_d$, $1000^2$ in output cost and $1000^3$ in training cost; compared to predicting the full state vector of a high-fidelity simulation.
In such a case, the training data generation via a high-fidelity simulation would be the most expensive step, as we assume in the paper, which would also mean that the RHS of \cref{eq:cost_analysis} would be much larger than the LHS.
Additionally, as remarked before, this approach also suits an experimental framework where running the experiment for a sufficiently long time might be expensive.
}

\bibliographystyle{apsrev4-2_nourl}
%\bibliography{biblio}%

\end{document}